\documentclass[referee,usenatbib,useAMS]{mn2e}
\usepackage{multirow}
\usepackage{graphicx}
\usepackage{times}
\usepackage{float}
\usepackage{pdflscape}
\usepackage{graphicx}
\usepackage{threeparttable}
\usepackage{epstopdf}
\usepackage{amsmath}
\usepackage{amssymb}
\usepackage{xspace}
\title[X-ray Flaring in PDS\,456]{X-ray flaring in PDS\,456 observed in a high-flux state}
\author[Matzeu et al.]
{G. A. Matzeu,$^{1}$\thanks{Correspondence to: g.matzeu@keele.ac.uk; g.a.matzeu@icloud.com} 
J. N. Reeves, $^{1,2}$ E. Nardini, $^1$ V. Braito, $^{2,3}$ T. J. Turner $^4$ and M. T. Costa $^1$ \\
$^1$Astrophysics Group, School of Physical and Geographical Sciences, Keele University, Keele, Staffordshire ST5 5BG, UK\\
$^2$Center for Space Science and Technology, University of Maryland Baltimore County, 1000 Hilltop Circle, Baltimore, MD 21250, USA\\
$^3$INAF – Osservatorio Astronomico di Brera, Via Bianchi 46, I-23807 Merate (LC), Italy\\
$^4$Department of Physics, University of Maryland Baltimore County, 1000 Hilltop Circle, Baltimore, MD 21250, USA}

\newcommand{\pds}{PDS\,456\xspace}
\newcommand{\xmm}{{\it XMM-Newton}\xspace}
\newcommand{\nustar}{\textit{NuSTAR}\xspace}

\newcommand{\suzaku}{{\it Suzaku}\xspace}

\newcommand{\Msun}{\hbox{$\rm\thinspace M_{\sun}$}}

\newcommand{\xspec}{\textsc{xspec\xspace}}

\newcommand{\optxagn}{\texttt{optxagnf}\xspace}

\begin{document}

\date{\today}

\pagerange{\pageref{firstpage}--\pageref{}} \pubyear{?}

\maketitle
\label{firstpage}


\maketitle
\begin{abstract}

\noindent We present an analysis of a $190$\,ks (net exposure) \textit{Suzaku} observation, carried out in 2007, of the nearby ($z=0.184$) luminous (L$_{\rm bol}\sim10^{47}$\,erg\,s$^{-1}$) quasar PDS\,456. In this observation, the intrinsically steep bare continuum is revealed compared to subsequent observations, carried out in 2011 and 2013, where the source is fainter, harder and more absorbed. We detected two pairs of prominent hard and soft flares, restricted to the first and second half of the observation respectively. The flares occur on timescales of the order of $\sim50$\,ks, which is equivalent to a light-crossing distance of $\sim10\,R_{\rm g}$ in PDS\,456. From the spectral variability observed during the flares, we find that the continuum changes appear to be dominated by two components: (i) a variable soft component ($<2$\,keV), which may be related to the Comptonized tail of the disc emission, and (ii) a variable hard power-law component ($>2$\,keV). The photon index of the latter power-law component appears to respond to changes in the soft band flux, increasing during the soft X-ray flares. Here the softening of the spectra, observed during the flares, may be due to Compton cooling of the disc corona induced by the increased soft X-ray photon seed flux. In contrast, we rule out partial covering absorption as the physical mechanism behind the observed short timescale spectral variability, as the timescales are likely too short to be accounted for by absorption variability.

\end{abstract}
\begin{keywords}
Subject headings: Black hole physics -- galaxies: active -- galaxies: nuclei -- quasars: individual (\pds) -- X-rays: galaxies 
\end{keywords}


\section{introduction}

It is widely accepted that the primary X-ray emission of AGN originates from the `seed' UV disc-photons, which then undergo inverse Compton scattering in a corona of relativistic electrons \citep{HaardtMaraschi91,HaardtMaraschi93}. These then produce the observed hard X-ray tail, usually phenomenologically described by means of a simple power law. The X-ray variability of AGN can be observed in both the intensity of the primary intrinsic continuum and the spectral shape \citep[e.g.][]{Green93}. Furthermore, principal component analysis of AGN has shown that the spectral variability can be often characterized by at least two components, a variable steep primary continuum and a less variable harder component \citep[e.g.,][]{Vaughan04,Miller07,Parker15}. Alternatively, the X-ray spectral variability has been explained through the presence of absorbing gas in the line-of-sight \citep{Risaliti02}, seen as changes in the covering fraction of a partial covering absorber \citep[e.g.][]{Turner11}. Moreover, partial covering scenarios have been successful in explaining the complex X-ray spectral properties of AGN in different energy bands such as: substantial continuum curvature below $10$\,keV \citep[e.g.,][]{Miller08,Miyakawa12}, rapid spectral variability \citep[e.g.,][]{NardiniRisaliti11} and pronounced hard excesses above $10$\,keV \citep[e.g.,][]{Tatum13}. 
\\
\indent The luminous quasar \pds, discovered by \citet{Torres97}, is located at redshift $z=0.184$ and has an observed bolometric luminosity of L$_{\rm {bol}}\sim10^{47}$ erg s$^{-1}$ \citep{Simpson99,Reeves00}, making it the most luminous quasar in the local Universe ($z<0.3$). Its high luminosity is more typical of quasars located at redshift $z=2$--$3$, the peak of the quasar epoch at which black hole feedback was thought to play a key role in the evolution of galaxies. Over the last decade, \pds has shown complex X-ray variability, most likely due to both absorption changes and variations in the intrinsic continuum \citep{Behar10}. The extreme X-ray nature of \pds was first noticed by \citet{Reeves00}, where very rapid X-ray variability, on timescales of $\sim15$\,ks, was observed in RXTE monitoring observations in the 3--10\,keV band. This indicates, by the light-crossing time argument, a very compact X-ray source of a few gravitational radii ($R_{\rm g}$) in extent\footnote{\pds has an estimated  black hole mass of $\sim10^{9}\,\Msun$, thus by the light-crossing time argument, a typical timescale of $\sim15$\,ks corresponds to a distance of $\sim3\,R_{\rm g}$} (where $R_{\rm g}=GM_{\rm BH}/c^2$). A short ($40$\,ks) observation carried out with \xmm in 2001 detected a strong absorption trough in the iron\,K band, above $7$\,keV, possibly due to the highly ionized iron\,K-shell feature with a corresponding outflow velocity of $v_{\rm w}\gtrsim 0.1c$ \citep{Reeves03}. A ($190$\,ks) \suzaku observation in 2007 firmly confirmed the evidence for this fast outflow, revealing two highly significant absorption lines centred at rest-frame energies of $E=9.09\pm0.05$\,keV and $E=9.64\pm0.08$\,keV, where no strong atomic transitions were otherwise expected. The association of these lines to the nearest expected strong line, the Fe\,\textsc{xxvi} Ly$\alpha$ transition at laboratory rest-energy of $E_{\rm lab,rest}=6.97$\,keV, implied an outflow velocity of $\sim0.25$--$0.30c$ \citep[][hereafter R09]{Reeves09}. Similarly in a more recent (2011) $120$\,ks \suzaku follow-up observation, a broader absorption trough (i.e., higher equivalent width) compared to the 2007 observation was again found at $E\sim9$\,keV (in the source rest frame), showing the variability of the iron\,K absorption feature. These observed changes could be, in fact, occurring within the same flow of gas in photo-ionization equilibrium with the emergent X-ray emission \citep[][hereafter R14]{Reeves14}. 
\\
\indent Following these important results, two large observational campaigns (each with $\sim500\,\rm ks$ net exposure) of \pds were performed in the period between 2013 and 2014. The first was with \suzaku in early 2013 and the second with \xmm and \nustar simultaneously in late 2013/early 2014. In the long ($\sim1$\,Ms duration) 2013 \suzaku campaign \pds was caught in an unusually low-flux state and it was possible to determine the timescales at which both the X-ray absorption and continuum variations occurred \citep[][M16 hereafter]{Gofford14,Matzeu16}. This was achieved by directly measuring the disc wind absorber's behaviour on timescales of hundreds of ks (M16), suggesting that the partial covering clouds could be the denser, or clumpy part of an inhomogeneous accretion disc wind. In this analysis, through the presence of a large X-ray flare, it was also possible to place an estimate on the radial extent of the X-ray emission region, which was of the order of $\sim$15--20\,R$_{\rm g}$, although the hard X-ray ($>2$\,keV) emission may have originated from a more compact or patchy corona of hot electrons, typically $\sim$6--8\,$R_{\rm g}$ in size. The second large campaign consisted in a series of five simultaneous observations with \xmm and \nustar in 2013--2014, where \citet[][N15 hereafter]{Nardini15} resolved a fast ($\sim0.25c$) P-Cygni like profile at Fe K, showing that the absorption originates from a wide angle accretion disc wind.    
\\
\indent In this work we present a detailed analysis of the 190\,ks (net exposure) \suzaku 2007 observation where the intrinsically bare continuum is revealed compared to the subsequent more absorbed \suzaku observations carried out in 2011 and 2013. The rapid variability that is exhibited by \pds during this observation, as well as the relatively clean determination of the primary continuum form, provides an opportunity to study the intrinsic variability mechanism in a high luminosity AGN. The exact origin of the rapid X-ray variability and coronal changes are not generally well understood in AGN at present, but may be related to magnetic flaring activity in the corona \citep[e.g.,][]{Leighly97,DiMatteo98,Merloni01,Reeves02}. For instance, \citet{Legg12} measured the different responses of the continuum over broad energy bands to short timescale flares seen in lower mass Narrow Line Seyfert 1 Galaxies, which were found to show contrasting and complex behaviour. In addition several authors have also parametrized the delayed responses between different energy bands in the form of lag spectra, which may be the result of reverberation delays on short timescales \citep[e.g.,][]{Zoghbi10,Kara13}, but whose exact interpretation is subject to debate \citep[e.g][]{Miller10}. In \pds, the high black hole mass is especially advantageous, as it makes it possible to directly measure the changes in the X-ray spectrum in response to flares over timescales of tens of ks, which in lower mass AGN are confined to very short timescales of $<1\,\rm ks$, precluding a direct spectral analysis.
\\
\indent Furthermore, it is also known that the broadband X-ray spectra from AGN consist of multiple continuum components, being dominated by a soft X-ray excess at energies below 1\,keV and a harder power-law like component that emerges at higher energies from a hot corona, possibly via thermal Comptonization \citep[e.g.,][]{HaardtMaraschi93}. The nature of this two component continuum is still under debate, particularly the form of the soft excess which has been accounted for via Comptonization of UV disc photons by a cooler population of electrons in a multi-temperature corona \citep[e.g][]{Done12}, but also in terms of relativistically blurred ionized reflection from the inner accretion disc \citep[e.g.,][]{Crummy06,Nardini11,Gallo13}. In these \pds observations, a further motivation is to study how these soft and hard continuum components evolve across a long observation, where pronounced short-term variability is present.
\\
\indent Here we focus on the broadband spectral variability of \pds on short timescales, as the 2007 \suzaku observation exhibits substantial short-term variability, which for brevity we will refer to as `flares' hereafter. In this observation several strong individual flares in both the soft and hard X-ray bands are detected. The previous results on this 2007 data set, which focused on the Fe K feature and fast ($\sim0.3c$) wind, were presented in R09. As we will show, the time-sliced spectra from \pds present an overall spectral evolution from hard to soft during the observation, which may be intrinsically associated with Comptonization in a compact corona of $\la10\,R_{\rm g}$ in extent. The paper is organized as follows; in Section\,\ref{sec:data_reduction} we summarize the data reduction, while in Section\,\ref{sec:broadband spectral analysis} we model the overall optical to hard X-ray SED during the high-flux 2007 Suzaku observation, utilizing contemporary \xmm \& \nustar observations at a similar flux level. In Section\,\ref{sec:Temporal Behaviour} we present an initial variability analysis and quantify the overall spectral evolution, while in Section\,\ref{subsec:Time Dependent Spectral Analysis} we present a detailed time-dependent spectral analysis. Section\,\ref{sec:Discussion} discusses the origin of the spectral variability, which is most likely to be the result of changes in the coronal X-ray emission of \pds.
\begin{table*}

\centering
\begin{tabular}{ccccccc}
\hline

Parameter                       &\suzaku 2007            &\suzaku 2013c          &\textit{XMM}/ObsB       &\textit{NuSTAR}/ObsB       \\
\hline

Obs.~ID                         &$701056010$             &$707035030$            &$0721010301$             &$60002032004$             \\

Start Date, Time (UT)           &2007-02-24, 17:58       &2013-03-08, 12:00      &2013-09-06, 03:24        &2013-09-06, 02:56         \\

End Date, Time (UT)             &2007-03-01, 00:51       &2013-03-11, 09:00      &2013-09-07, 10:36        &2013-09-07, 10:51         \\

Duration(ks)                    & 370                    & 248.401               &112.3                    &113.9                     \\

Exposure(ks)$^{\rm a}$              & 190.6                  & 108.3                 &92.2                     &43.0                      \\

Flux$_{(0.5-2)\rm keV}^{b}$     & 3.46                   & 0.43                  &2.6                      & --                       \\

Flux$_{(2-10 )\rm keV}^{b}$     &3.55                    & 1.72                  &2.8                      &2.9                       \\

Flux$_{(15-50)\rm keV}^{b}$     & $5.7\pm2.2$            &$<1.97$                & --                      &0.89                      \\

\hline

\end{tabular}
\caption{Summary of the 2007 and the third 2013 (2013c) \suzaku observations of \pds plus the second 2013 (ObsB) \xmm \& \nustar observation for comparison purposes, which is similar in spectral form to the 2007 \suzaku observation.}
\vspace{-5mm}
\begin{threeparttable}
\begin{tablenotes}
	\item[a] Net Exposure time, after background screening and dead-time correction.
	\item[b] Observed fluxes in the 0.5--2\,keV, 2--10\,keV and 15--50\,keV bands in units $\times10^{-12}$\,erg cm$^{-2}$ s$^{-1}$ ($1\sigma$ upper limits).	
\end{tablenotes}
\end{threeparttable}
\label{tab:table_1}

\end{table*}

\section{Data reduction}
\label{sec:data_reduction}

\suzaku \citep{Mitsuda07} observed \pds between the 24th of February and the 1st of March 2007 through the X-ray Imaging Spectrometer \citep[XIS;][]{Koyama07} and the Hard X-ray Detector \citep[HXD;][]{Takahashi07}. In this observation \pds was marginally detected in the HXD/PIN, although with large uncertainties in the hard X-ray flux due to background systematics (R09), while there was no detection in the HXD/GSO detector. The observation is constituted by a single continuous sequence (see Table\,\ref{tab:table_1} OBSID:701056010, hereafter 2007), with a duration of $\sim370$ ks. The details for the 2007 observation are listed in Table\,\ref{tab:table_1}. R09 found that the front illuminated XIS-FI CCDs (XIS 0, XIS 3) and the back illuminated XIS-BI XIS 1 spectra are consistent with each other; however in this work we focus, for simplicity, on the XIS-FI spectra. They were subsequently merged into a single XIS-FI 03 spectrum, where for the combined spectrum we adopted a minimum grouping of 50 counts per spectral bin in order to adopt the $\chi^2$ minimisation technique. For all subsequent spectral fitting, we consider the XIS-FI spectrum between 0.6--10\,keV and we remove the 1.7--2.1\,keV band to avoid uncertainties associated with the detector Si\,K edge.
Values of H$_0=70$ km s$^{-1}$ Mpc$^{-1}$ and $\Omega_{\Lambda_{0}}=0.73$ are assumed throughout and errors are quoted at $90\%$ confidence level ($\Delta \chi^{2}=2.71$), for one parameter of interest.

\begin{figure*}
\
\centering	
\includegraphics[scale=0.70]{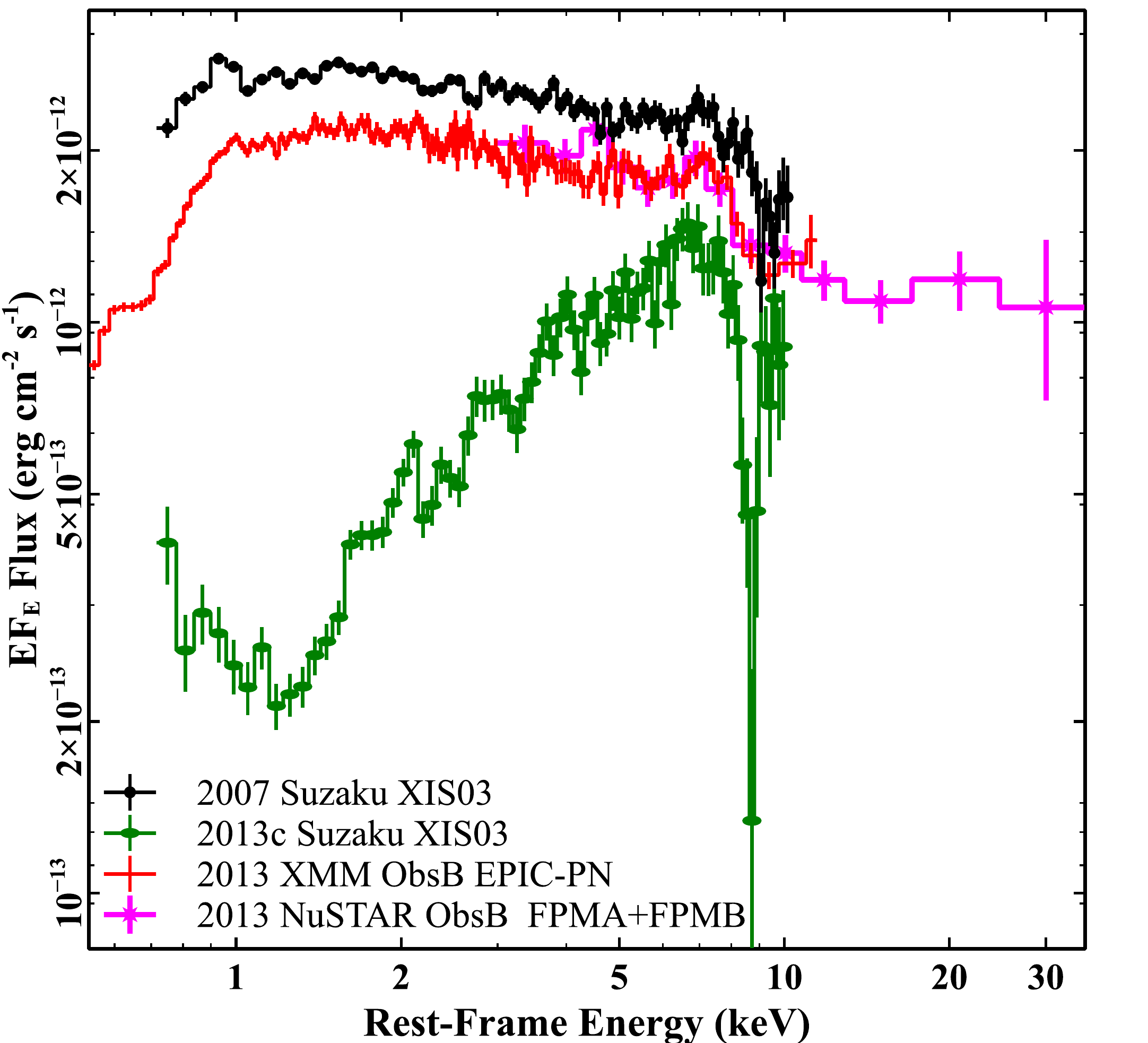}	
\caption{XIS spectra of the 2007 \suzaku observation (black). For comparison, the 2013 ObsB spectra of the \xmm \& \nustar campaign (red and magenta respectively), together with the 2013c (green) \suzaku spectrum, are also shown. Through a visual inspection, the 2007 and ObsB spectra are remarkably similar in shape. Overall, the long-term spectral variations in the soft band and in the iron-K region, which characterize this type 1 quasar, are conspicuous. Note that the spectrum becomes harder and more absorbed during the lower flux 2013 \suzaku observation. The spectra were unfolded against a simple $\Gamma=2$ power law and are not corrected for Galactic absorption.}
\label{fig:pds456_suz_xmm_nu_spectra}
\end{figure*}

\section{Broadband Spectral Analysis}
\label{sec:broadband spectral analysis}

All the spectral analysis and model fitting was carried out with \xspec\,v12.9.0 \citep{Arnaud96}. Initially we compare the \suzaku 2007 spectrum to some of the more recent \xmm \& \nustar data that captured the source in a similar flux state. More specifically, Fig.\,\ref{fig:pds456_suz_xmm_nu_spectra} shows this comparison where the 2007 \suzaku observation and the second sequence (ObsB, hereafter) from the simultaneous \xmm \& \nustar campaign, carried out in late 2013 (see N15 for details), are shown in black and red respectively. Although the \xmm \& \nustar campaign consists of five sequences, the second observation is very similar in flux and photon index to the 2007 \suzaku observation, as measured in the 2--8\,keV rest-frame band (e.g., \xmm \& \nustar ObsB $\Gamma_{2-8}=2.24\pm0.03$ versus \suzaku, $\Gamma_{2-8}=2.25\pm0.03$). Indeed the only difference is a slight offset of $\sim20\%$ in flux, due to the \suzaku observation being slightly brighter. In Fig.\,\ref{fig:pds456_suz_xmm_nu_spectra} we also show for comparison the last \suzaku sequence from the large campaign consisting of three observations in 2013, carried out approximately six months prior to the \xmm \& \nustar observation. In contrast to the 2007 \suzaku observation, in 2013 \suzaku caught \pds at a considerably lower flux, where the third sequence (2013c hereafter) has the lowest flux state to date and is much harder in shape (see M16 for more details). The relative fluxes for these 2007 and 2013c \suzaku observations, as well as the \xmm \& \nustar (ObsB) observation are also summarized in Table\,\ref{tab:table_1}.
\\
\indent During the 2007 \suzaku observation (R09), \pds was observed with little obscuration, which revealed a steep intrinsic continuum ($\Gamma>2$) as evident in Fig.\,\ref{fig:pds456_suz_xmm_nu_spectra}. The comparable characteristics between the 2007 \suzaku and ObsB spectra are also illustrated in Fig.\,\ref{fig:pds456_suz_xmm_nu_spectra}. Note that as the HXD detection gives a large uncertainty on the hard X-ray ($>10\,\rm keV$) flux, due to the large systematics of the HXD/PIN background subtraction (R09), we adopted the \nustar data instead in order to characterize the broadband spectrum. The imaging characteristics of the FPMA and FPMB detectors provide a much more reliable measurement of the spectrum above 10\,keV. Thus, the \nustar observations do not show any evidence of a strong hard X-ray excess (also see N15) consistent with a simple extrapolation of the steep power-law to high energies.

\begin{table*}

\begin{tabular}{cc|c|}

\hline
Component                                &Parameter                                 &\suzaku XIS03 2007 \\
\hline

\multirow{1}{*}{\texttt{Tbabs}$^{*}$}    &$N_{\textsc{h},\rm{Gal}}/{\rm cm^{-2}}$                 & $2.0\times10^{21}$                     \\

\\

\multirow{6}{*}{\texttt{optxagnf}}       &$\log(L/L_{\rm Edd})$                     &$-0.13_{-0.12}^{+0.11}$\\

                                         &r$_{\rm cor}$ (R$_{\rm g}$)               &$38_{-28}^{+43}$\\

                                         &kT$_{\rm e}$ (eV)                         &$292_{-20}^{+32}$\\

										&$\tau$                                    &$12.4_{-1.4}^{+1.2}$\\

										&$\Gamma$                                  &$2.31_{-0.04}^{+0.06}$\\

										&F$_{\rm pl}$                              &$0.08_{-0.03}^{+0.02}$\\

\\

\multirow{2}{*}{\texttt{pc}}

                                        &$\log$($N_{\rm{H,low}}$/cm$^{-2}$)         &$23.1_{-0.3}^{+0.1}$\\

                                        &f$_{\rm cov}$                              &$0.15\pm0.07$\\

\\

\multirow{4}{*}{Gaussian$_{\rm em,Fe\,K}$}     

										&Energy (keV)$^{\rm a}$                    &$7.07_{-0.14}^{+0.15}$\\

                                         &norm$^b$                                  &$3.89\pm1.84$\\

                                         &$\sigma$ (eV)                             &$260_{-118}^{+239}$\\

										&EW (eV)                                   &$88_{-42}^{+41}$\\


\\

\multirow{4}{*}{Gaussian$_{\rm abs,Fe\,K}1$}

                                         &Energy (keV)$^{\rm a}$                    &$9.07_{-0.06}^{+0.05}$\\

                                       	&norm$^b$                                  &$-2.48_{-0.93}^{+1.21}$\\

                                         &$\sigma$ (eV)                             &$<160$\\

										&EW (eV)                                   &$105_{-40}^{+51}$	\\


\\

\multirow{4}{*}{Gaussian$_{\rm abs,Fe\,K}2$}

                                         &Energy (keV)$^{\rm a}$                    &$9.54_{-0.08}^{+0.09}$\\

                                         &norm$^b$                                  &$-1.88_{-1.03}^{+0.89}$\\

                                         &$\sigma$ (eV)                             &$<160^{\rm t}$\\

										&EW (eV)                                   &$94_{-52}^{+45}$\\


\\

\multirow{4}{*}{Gaussian$_{\rm em,soft}\,1$}

                                        &Energy (keV)$^{\rm a}$                     &$0.93_{-0.03}^{+0.02}$\\

                                        &norm$^b$                                   &$125.7_{-54.6}^{+101.5}$\\

                                        &$\sigma\rm\,(eV)$                          &$45_{-25}^{+27}$\\

                                        &EW (eV)                                    &$24^{+8}_{-7}$             \\
                                        

\\

\multirow{4}{*}{Gaussian$_{\rm em,soft}\,2$}

                                       &Energy (keV)$^{\rm a}$                      &$1.15\pm0.02$                \\

                                       &norm$^b$                                    &$18.8_{-1.0}^{+0.9}$                \\

                                       &$\sigma\rm\,(eV)$                           &$10^*$                       \\

                                       &EW (eV)                                     &$7^{+4}_{-3}$               \\
                                       

\\
&~~~~~~~~~~~~~~$\chi^{2}_{\nu}=910/802$&\\

\hline
\end{tabular}

\caption{$^{\rm t}$ denotes that the parameter is tied during fitting, $^*$ indicates a parameter fixed during fitting. Note that as the \suzaku and \nustar spectra are not simultaneous we accounted for any hard X-ray flux differences using a cross-normalization constant factor during fitting, of $0.80\pm0.10$ for \nustar ObsB compared to \suzaku 2007.}
\vspace{-5mm}
\begin{threeparttable}
\begin{tablenotes} 
	\item[\small{L/L$_{\rm Edd}$: Eddington ratio,}]  
	\item[\small{r$_{\rm cor}$ radius of the X-ray corona in R$_{\rm g}$,}]
	\item[\small{F$_{\rm pl}$: fraction of the dissipated accretion energy emitted in the hard power-law,}]	
 	\item[\small{\texttt{pc}: partial covering component with respective column density and covering fraction,}] 
	\item[a] rest-frame energy of the emission and absorption Gaussian profiles,
	\item[b] Gaussian emission and absorption profile normalization, in units of $10^{-6}$ photons cm$^{-2}$ s$^{-1}$.
Note that in this work, for simplicity, the spin parameter \texttt{a$^{\star}$} was kept fixed to zero in all the \texttt{optxagnf} fits. 
\end{tablenotes}
\end{threeparttable}
\label{tab:sed}
\end{table*}

\subsection{Modelling The Broadband SED}
\label{subsec:Modelling The Broadband SED}

\begin{figure}
\centering	
\includegraphics[scale=0.80]{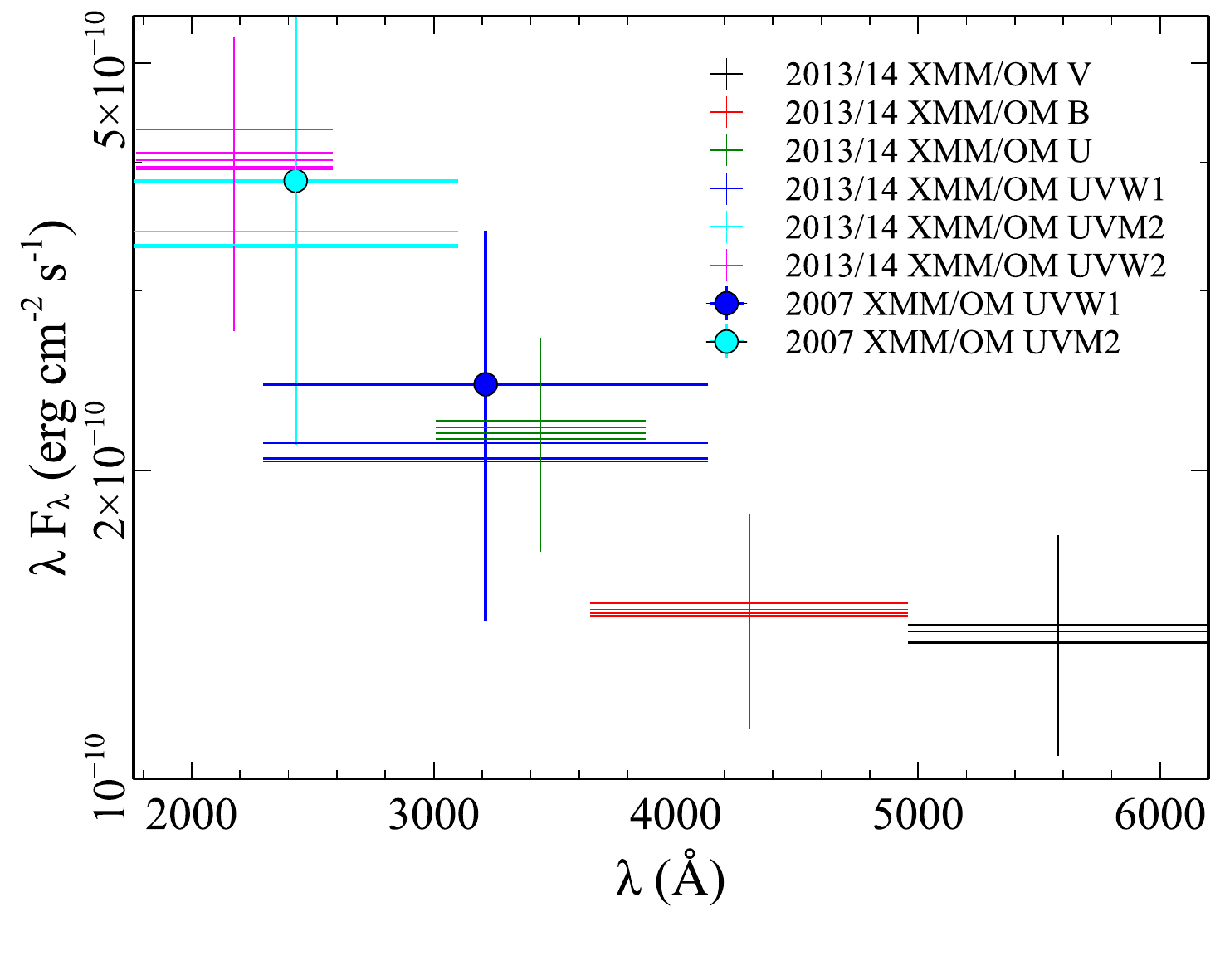}	
\caption{The \xmm OM photometric data of \pds taken between 2007 and 2014. Overall only small changes are observed in the optical/UV band, typically of the order of $5$--$20\%$ and are consistent within the uncertainties. Thus, when constructing the optical to hard X-ray SED, we can adopt the OM data from ObsB regardless of the fact that they are non simultaneous with the 2007 \suzaku observation.}
\label{fig:pds_2001_07_13_14_OM}
\end{figure}

\begin{figure}
\centering	
\includegraphics[scale=0.70]{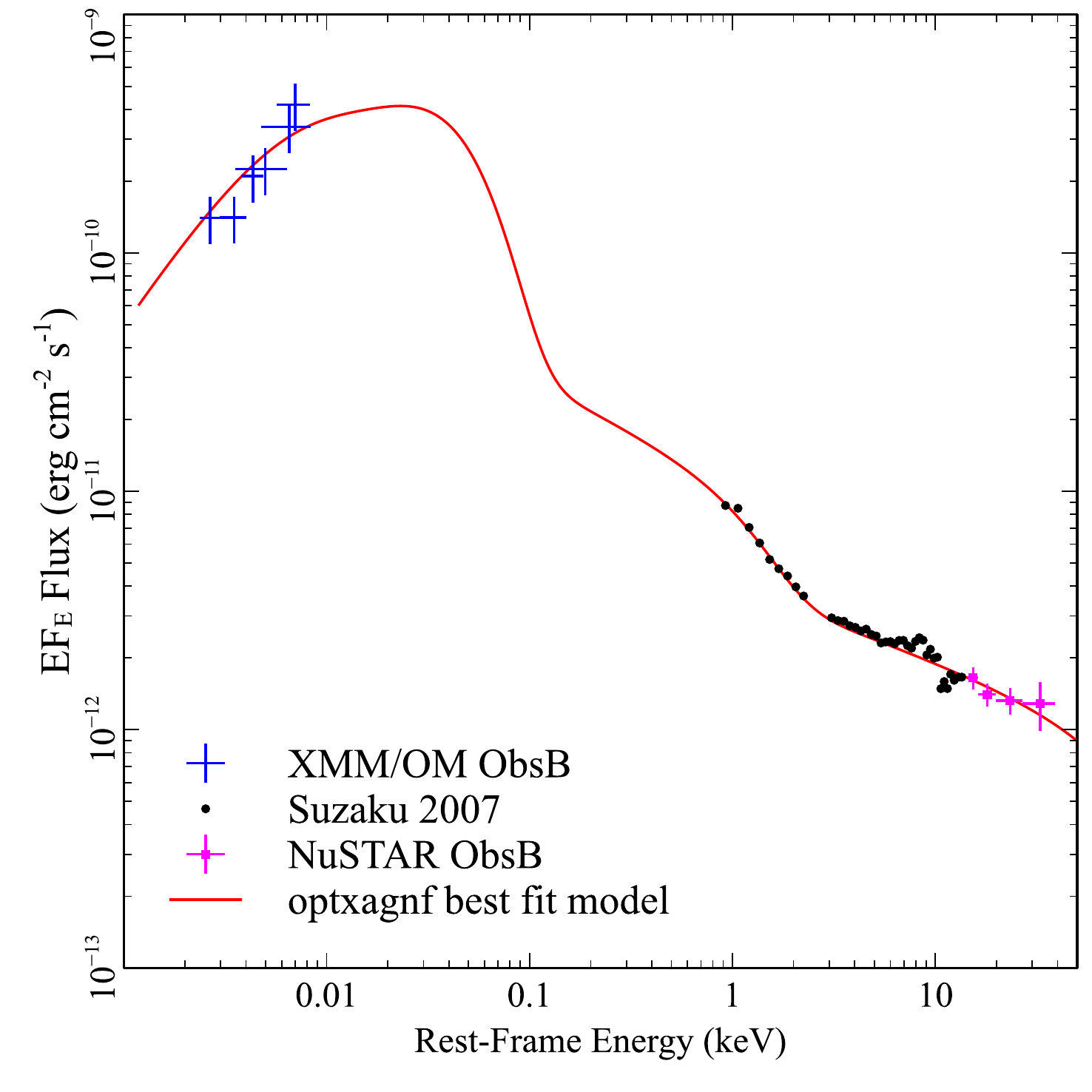}	
\caption{Optical to hard X-ray spectral energy distribution (SED) of \pds obtained by combining the OM and \nustar data (blue and magenta respectively) from the \textit{XMM-Newton/NuSTAR} observation in 2013 (ObsB) and the 2007 time-averaged spectrum from the \suzaku campaign (black). For clarity, the NuSTAR data is only plotted above 10\,keV. The SED was fitted with the \texttt{optxagnf} model (red solid line). Note that the model is corrected for absorption (either Galactic or local to the source), the OM data has been dereddened as is described in Section\,\ref{subsec:Modelling The Broadband SED} while the X-ray data have been corrected for a Galactic absorption column of $2.0\times10^{21}\,{\rm cm}^{-2}$.}
\label{fig:pds_2007_optxagnf_model}
\end{figure}

\begin{figure}
\centering	
\includegraphics[scale=1.0]{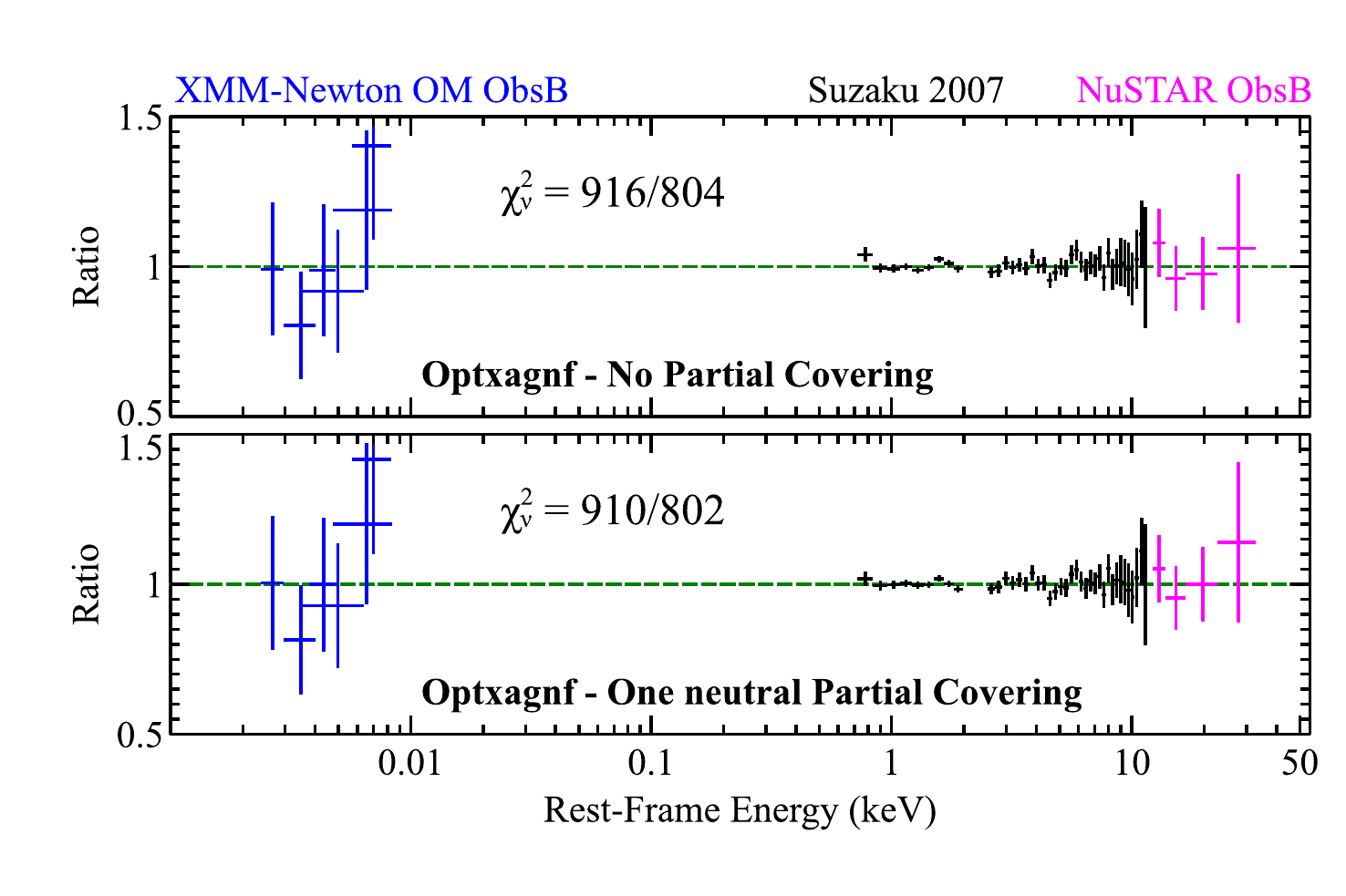}	
\caption{Residuals for the different \texttt{optxagnf} model fits over the $1\,\rm eV$--$50\,\rm keV$ rest-frame energy range, plotted as data/model ratios. The spectra correspond to the \xmm OM (blue) and \nustar from ObsB in 2013 (magenta) respectively, while also plotted is the \suzaku 2007 (black) time-averaged spectrum with the Galactic absorption included. The various Gaussian emission and absorption lines, as described in Section\,\ref{subsec:Modelling The Broadband SED}, have been included in the model. Top panel: SED fits with \texttt{optxagnf} and no partial covering absorption. Bottom panel: Same as above but with the addition of one neutral partial covering layer. There is no apparent difference between the two cases, with little intrinsic absorption required in the high-flux data. For extra clarity, the \suzaku and \nustar spectra have been re-binned by a factor of 4 and 8 respectively.}
\label{fig:sed_ra_test}
\end{figure}

We now attempt to characterize the broadband spectral energy distribution (SED) of \pds in this high-flux (unabsorbed) state. In particular, we tested whether the broadband SED could be described by a multi-temperature Comptonized accretion disc model, using the \texttt{optxagnf} model \citep{Done12} in XSPEC. This model is characterized by three distinct components, which are self-consistently powered by dissipation in the accretion flow: (i) the thermal emission from the outer accretion disc in the optical/UV; (ii) the Comptonization of the UV disc photons into a soft X-ray excess from a `warm' disc atmosphere; (iii) a high temperature Comptonization component from the `hot' corona (i.e. the standard hard X-ray power-law continuum). The parameter r$_{\rm cor}$ is the coronal size that acts as a transitional radius from the colour temperature corrected blackbody emission (produced from the outer disc), to a hard power law produced through inverse Compton scattering. The parameter \texttt{F$_{\rm pl}$} quantifies the fraction of the energy released in the power-law component. The parameters $kT$ and $\tau$ are the electron temperature and the optical depth of the soft Comptonization component respectively, possibly originating from the warm disc atmosphere and seen as the soft excess, while $\log({\rm L}/{\rm L_{Edd}})$ is the Eddington ratio of the AGN (see \citealt{Done12} for more details). In this work, for simplicity, the spin parameter \texttt{a$^{\star}$} was kept fixed at zero in all the \texttt{optxagnf} fits, noting that the subsequent spectral parameters are not strongly dependent on the black hole spin.
\\ 
\indent The form of the optical/UV to hard X-ray SED has been defined by using a combination of both \xmm \& \nustar ObsB spectra, as previously shown in Fig.\,\ref{fig:pds456_suz_xmm_nu_spectra}, which are in a similar form in the X-ray band to the \suzaku 2007 spectrum. Thus we adopted the optical/UV \xmm optical monitor (OM) and the \nustar data (above $10$\,keV) associated with ObsB sequence together with the \suzaku data. We also included a cross-normalization constant between the \nustar and \suzaku spectra, to allow for a small offset in flux, which was found to be $0.80\pm0.10$. It is important to note that when we make use of the optical/UV photometry, although the optical/UV fluxes are not simultaneous, there appears to be little variability in this band as observed from all the \xmm OM observations from 2007 to 2014 inclusive (to $\la20\%$), as shown in Fig.\,\ref{fig:pds_2001_07_13_14_OM}. 
\\
\indent In order to fit the SED with the \texttt{optxagnf} model, we included the effect of Galactic absorption in the X-ray band and also corrected the OM data points for reddening based on the standard \citet{Cardelli89} extinction law, with A(V) and E(B--V) values relevant to the source \citep{Simpson99,O'Brien05}. Overall, the \texttt{optxagnf} model reproduced well the shape of the UV bump, the soft excess and the hard X-ray power-law component. The overall SED and best fit \texttt{optxagnf} model are shown in Fig.\,\ref{fig:pds_2007_optxagnf_model}. We note that, in contrast to the 2013 \suzaku campaign where \pds appeared highly absorbed (see M16), there is no prominent spectral curvature above $1$\,keV (see Fig.\,\ref{fig:sed_ra_test} top panel). Thus, it was possible to achieve an acceptable fit, with $\chi^{2}/\nu=916/804$, without the inclusion of any absorbers. Statistically speaking, adding some absorption to the broadband SED leads to a very marginal improvement, $\Delta\chi^{2}/\Delta\nu=6/2$ to the fit, shown in Fig.\,\ref{fig:sed_ra_test} (bottom panel). Thus in 2007, the degree of absorption is minimal when parametrized with one layer of a neutral partial coverer (\texttt{zpcfabs}) of column density $\log(N_{\rm H}/{\rm cm^{-2}})=23.1_{-0.3}^{+0.1}$, as then the covering fraction is very low with $f_{cov}=0.15\pm0.07$.
\indent Nonetheless, the \texttt{optxagnf} parameters of the 2007 data are similar to what was found in the highly absorbed 2013 \suzaku spectrum, once the latter is corrected for absorption. In the 2007 data, the coronal size is found to be $r_{\rm cor}=38_{-28}^{+43}\,R_{\rm g}$ and the optical depth is $\tau=12.4_{-1.4}^{+1.2}$. The temperature of the warm electrons, responsible for the Comptonized soft excess, is $kT=292_{-20}^{+32}$\,eV, whilst the fraction of energy released in the power-law is found to be F$_{\rm pl}=0.08_{-0.03}^{+0.02}$. The Eddington ratio of $\log({\rm L}/{\rm L_{Edd}})=-0.13_{-0.12}^{+0.11}$ implies that \pds radiates close to its Eddington luminosity ($\sim80\%$ of $L_{\rm Edd}$). This is consistent with what was found in M16 and with the expectation of a high accretion rate for \pds, given its black hole mass ($\sim10^{9}\,\Msun$) and high bolometric luminosity of $L_{\rm bol}\sim10^{47}$\,erg s$^{-1}$ \citep{Simpson99,Reeves00}. The X-ray spectral shape is also consistent with what was found in R09 and M16 with a best fit photon index of $\Gamma=2.31_{-0.04}^{+0.06}$ (see Table\,\ref{tab:sed} for more details). Thus, the high-flux 2007 SED of \pds reveals the bare AGN continuum, which can be consistently fitted by disc/coronal models.

\begin{figure*}
\centering	
\includegraphics[scale=0.5]{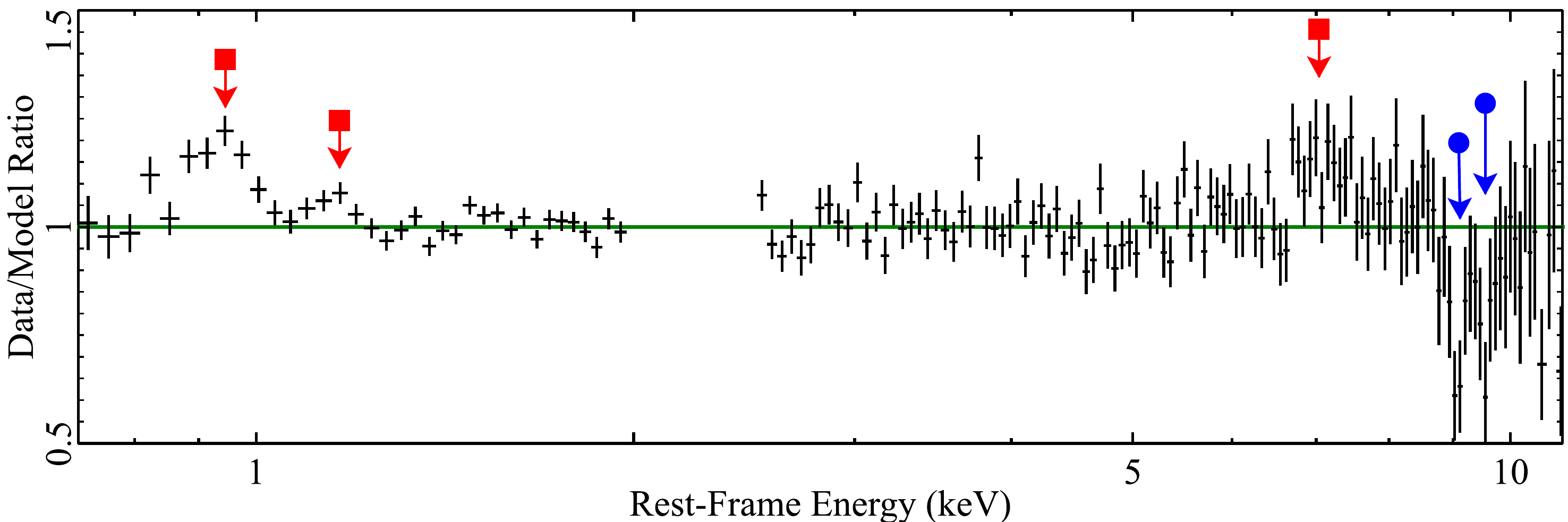}	
\caption{Data to model ratio, compared to the best-fit \optxagn model, plotted for the time-averaged \suzaku 2007 XIS03 spectrum between the rest-frame energy of 0.7--11\,keV showing the residual emission and absorption lines where their positions are indicated by the square/red and circle/blue arrows respectively. The positions of these lines features are consistent with those originally reported in R09.}
\label{fig:pds456_2007_time_spectum_new}
\end{figure*}

\indent In R09, the iron\,K absorption profile was resolved into two components at $E\sim9.1$\,keV and $E\sim9.6$\,keV that could be associated with He and H -like iron with two offset velocities. Here, as we changed the parametrization of the underlying continuum, we checked the consistency of these absorption features. Following the same approach as in R09, the iron\,K absorption is fitted with two (cosmologically redshifted) Gaussian components at rest-frame energies of $E=9.07_{-0.06}^{+0.05}$\,keV and $E=9.54_{-0.08}^{+0.09}$\,keV. As per R09, we also adopted three further Gaussian components to parameterise the ionized emission profile at $7.07_{-0.14}^{+0.15}$\,keV (in the quasar rest frame), and two soft X-ray emission lines with the stronger at $E=0.93_{-0.03}^{+0.02}$\,keV and the weaker at $E=1.15\pm0.02$\,keV. These five lines are consistent with what was found in R09 and R14 and for completeness we tabulate them in Table\,\ref{tab:sed}. The residuals of the emission (square/red arrows) and absorption (circle/blue arrows) lines are shown in Fig.\,\ref{fig:pds456_2007_time_spectum_new} where the \suzaku 2007 time-averaged spectrum was plotted as a ratio to the \optxagn model. As discussed in N15 and more recently in \citet{Reeves16}, the emission lines may represent the re-emission that occurs from the outflow in \pds, as seen in the Fe\,K band and measured in the soft X-ray grating spectra with the \xmm reflection grating spectrometer (RGS).
\\
\indent Indeed \citet{Reeves16} identified the above soft X-ray line features with a blend of recombination emission from primarily O\,\textsc{viii}, Ne\,\textsc{ix} and Ne\,\textsc{x}. These emission lines were especially prominent in the lower flux observations when the AGN was more absorbed and also when the broad absorption profiles from the wind were revealed near to 1\,keV.

\begin{figure}
\begin{center}
\includegraphics[scale=0.55]{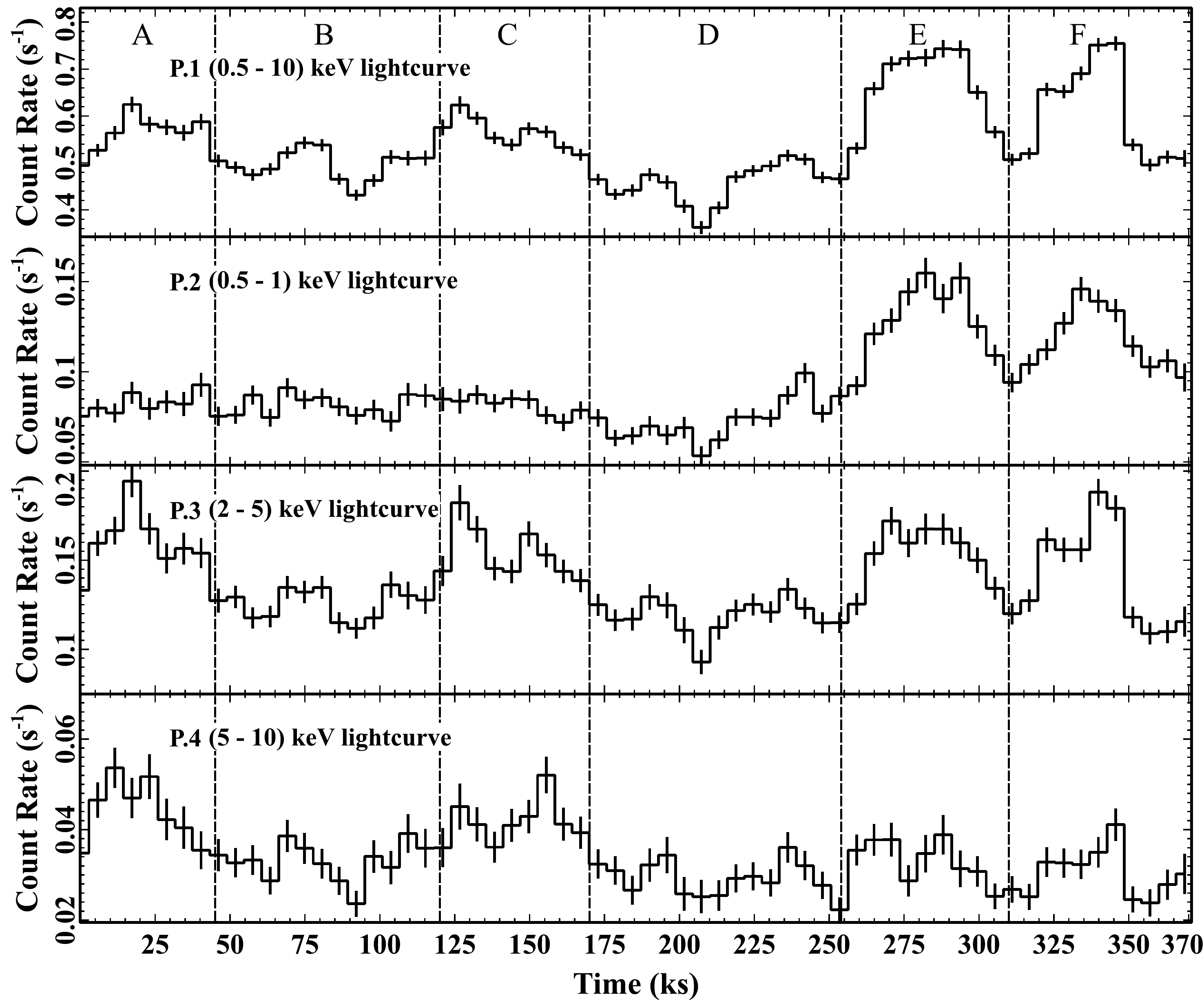}	
\caption{Plots showing the light curves extracted in different energy bands revealing some distinct behaviours. The two extremes are seen in the in the soft band (0.5--1\,keV) and in the hard band (5--10\,keV) where in the former we see two prominent flares only detected towards the end of the observation, whereas in the latter we observe two distinct flares confined to the first half of the observation. In the broadband (0.5--10\,keV) and in particular the middle band (2--5\,keV) we observe a blend of the two events (see text for more details). The bin size of the light curves are 5760 s, corresponding to one satellite orbit.}
\label{fig:pds_2007_3lc_hr}
\end{center}
\end{figure}

\section{Temporal Behaviour}
\label{sec:Temporal Behaviour}

Having parametrized the broadband 2007 high-flux spectrum of \pds, here our primary aim is to characterize the variability of \pds during this observation. Initially, we extracted the light curves corresponding to different energy bands, and their behaviour was subsequently compared.

\begin{figure}
\centering	
\includegraphics[scale=0.55]{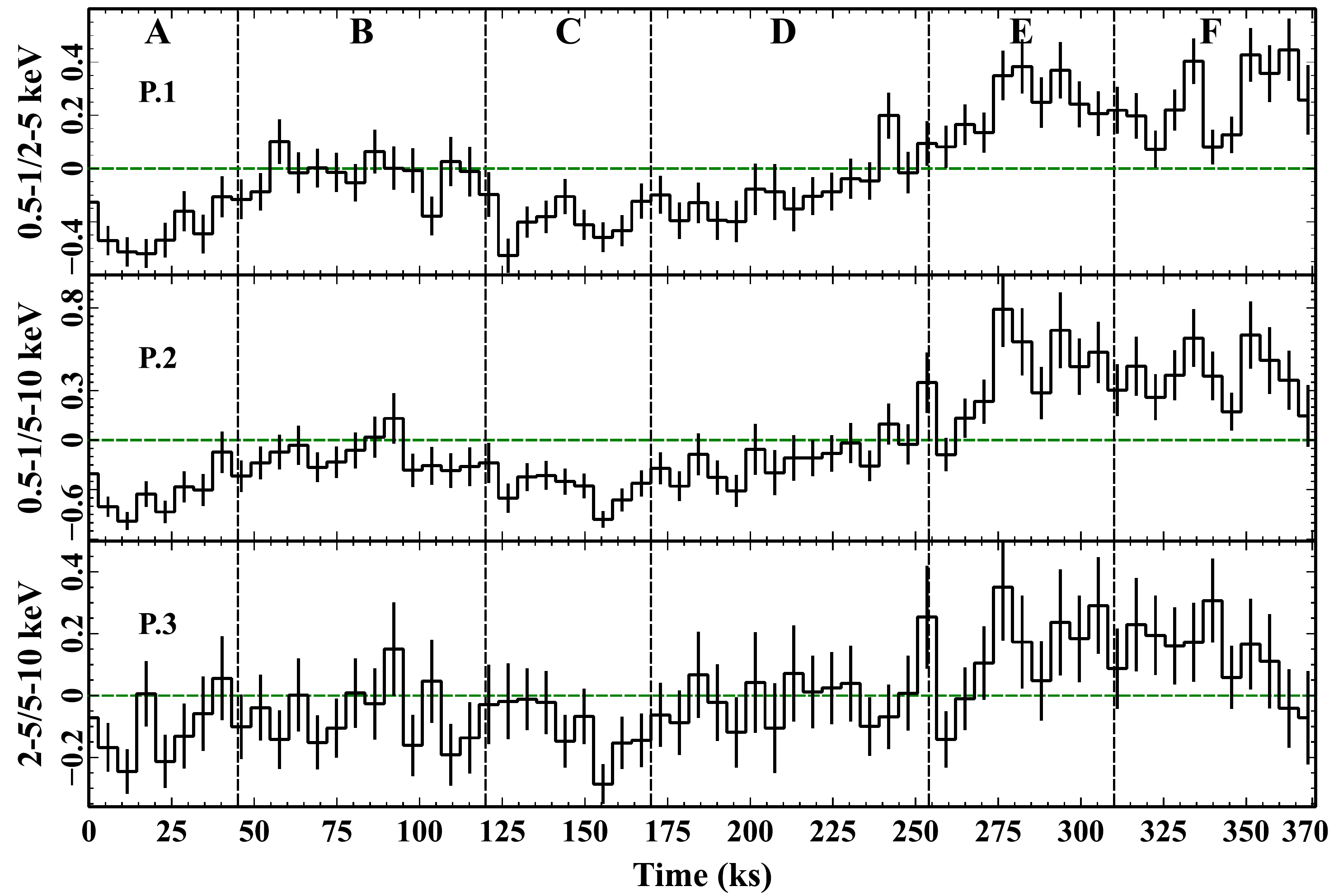}	
\caption{The normalised softness ratios (NSR) defined mathematically as ${\rm NSR}=\frac{{\rm SR(t)}-\left \langle {\rm SR} \right \rangle}{\left \langle {\rm SR} \right \rangle}$, where SR is the softness ratio between the respective energy bands. Note that positive and negative values correspond to the respective softening and hardening of the source with respect to the average spectral state. The dashed vertical lines identify the boundaries of the six individual slices.}    
\label{fig:pds_2007_3sr}
\end{figure}

\subsection{Description of the light curves and Softness Ratios}
\label{Description of the light curves and Softness Ratios}

In Fig.\,\ref{fig:pds_2007_3lc_hr} (Panel 1) we plot the overall broadband light curve between 0.5--10\,keV of the 2007 \suzaku observation of \pds, and according to the overall behaviour, we defined six distinct slices. Slice A, between 0--45\,ks, corresponds to a first minor flare, which is followed by a quiescent period, between 45--120\,ks, denoted as slice B. Between 120--170\,ks into the observation, a second minor flare is detected (slice C), followed again by a quiescent state (slice D), between 170--255\,ks, where the observation reaches its lowest flux at $\sim210$\,ks. Towards the end of the observation, a first major flare of $\sim55$\,ks in duration is detected between 255--310\,ks (slice E), followed by a second major flare of $\sim60$\,ks, between 310--370\,ks (slice F). Moving to narrower energy bands, the 0.5--1\,keV soft band light curve plotted in Fig.\,\ref{fig:pds_2007_3lc_hr} (P.2) is completely dominated by two flares at the end of the observation; whereas in the 5--10\,keV hard band light curve (Fig.\,\ref{fig:pds_2007_3lc_hr} P.4) we detect two prominent flares only in the first half of the observation. Thus the soft band flares (slices E and F) are not present at all in the hardest 5--10\,keV band and likewise, the hard band flares (slices A and C) are not present in the softest 0.5--1\,keV band. On this basis, we can differentiate between two soft and two hard events across the entire observation. Interestingly in the 2--5\,keV band light curve (Fig.\,\ref{fig:pds_2007_3lc_hr} P.3) we observe a blend of these separate (soft and hard) events. This indicates a two component behaviour with distinct variability in the soft and hard bands, the physical implications of which will be discussed later in Section\,\ref{net_flares}.   
\\
\indent After extracting the light curves in the above energy bands, we computed the corresponding normalised softness ratios (NSR hereafter) defined mathematically as ${\rm NSR}=\frac{{\rm SR(t)}-\left \langle {\rm SR} \right \rangle}{\left \langle {\rm SR} \right \rangle}$, showing the fractional change in the softness ratio between two energy bands when compared to the mean value. The 0.5--1\,/2--5\,keV, the 0.5--1\,/5--10\, keV and the 2--5\,/5--10\,keV NSRs are plotted in Fig.\,\ref{fig:pds_2007_3sr}. By inspecting all the three NSRs it is revealed, particularly in Fig.\,\ref{fig:pds_2007_3sr} (P.1) and (P.2), that in the first half of the observation the source is dominated by hard photon counts. On the other hand, at $\sim170$\,ks into the observation, the source becomes progressively softer, reaching its peak towards the end. Thus these three NSR plots confirm that the hardening of the spectrum coincides with the hard events, seen earlier in Fig.\,\ref{fig:pds_2007_3lc_hr} (P.4), and also indicate that the gradual softening of the source starts before the soft flares are detected (see Fig.\,\ref{fig:pds_2007_3lc_hr} P.2).

\subsubsection{Timescale Of The Flares}
\label{subsec:Properties Of The Flares}

In order to analyse the flares observed in both the hard and soft bands, we focused on the corresponding 0.5--1\,keV and 5--10\,keV light curves. As described above, during the latter part of the observation we detect an increase of the soft X-ray flux by a factor of $\sim3.5$, starting at $\sim200$\,ks into the observation and peaking during the flare in slice E over a duration of $\Delta t_{\rm soft}\sim90$\,ks. On the other hand, the hard X-ray flux shows an increase by a factor $\sim2.5$ between $\sim90$--160\,ks ($\Delta t_{\rm hard}\sim70$\,ks).
\\
\indent The doubling time for both the soft and the hard band is $t_{\rm double}\sim50$\,ks. This corresponds to an X-ray emission region size of about $\sim10\,R_{\rm g}$ for \pds. Note that this is also consistent with the earlier \texttt{optxagnf} disc plus corona model representation, where the lower limit of the coronal size was found to be $r_{\rm cor}\ga10\,R_{\rm g}$ (see Section\,\ref{subsec:Modelling The Broadband SED}).

\subsection{Flux-Flux Analysis}
\label{subsec:Flux-Flux plot}

\begin{figure}
\centering	
                                                                                                                                                                                                                                                                                                                                                                                                                                                                                                                                                                                                                                                                                   \includegraphics[scale=0.85]{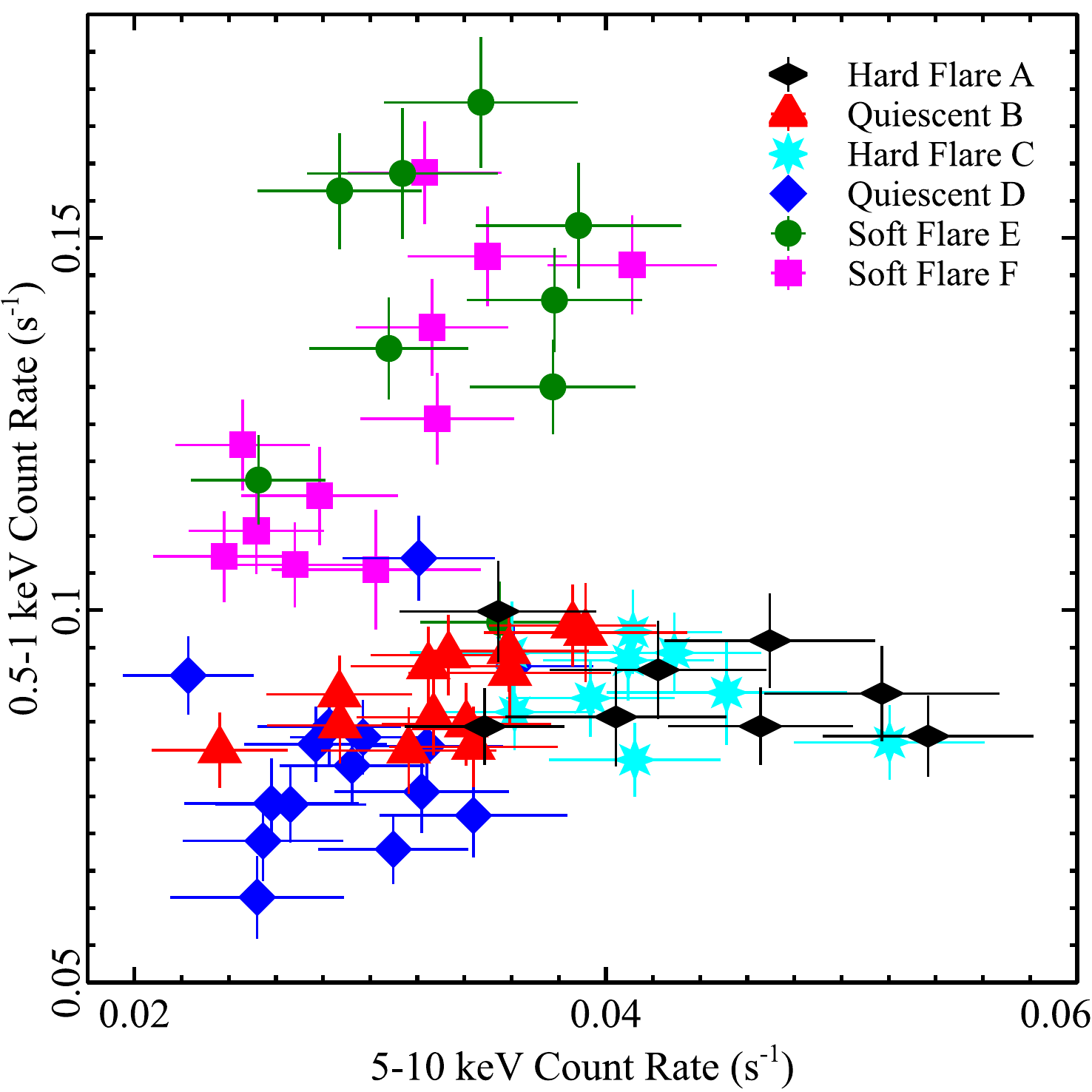}	
\caption{Flux-Flux plot for \pds between 5--10\,keV and 0.5--1\,keV where the distributions of the individual six slices are clearly labelled. The hard flare (A + C), soft flare (E + F) and quiescent (B + D) slices fall on distinct portions of the flux-flux plot, with little overlap between them. Each point on the plot corresponds to one orbital bin (5760\,s) during the observation.}    
\label{fig:pds_2007_count_count_6slice}
\end{figure}

\begin{figure}
\centering	
\includegraphics[scale=0.60]{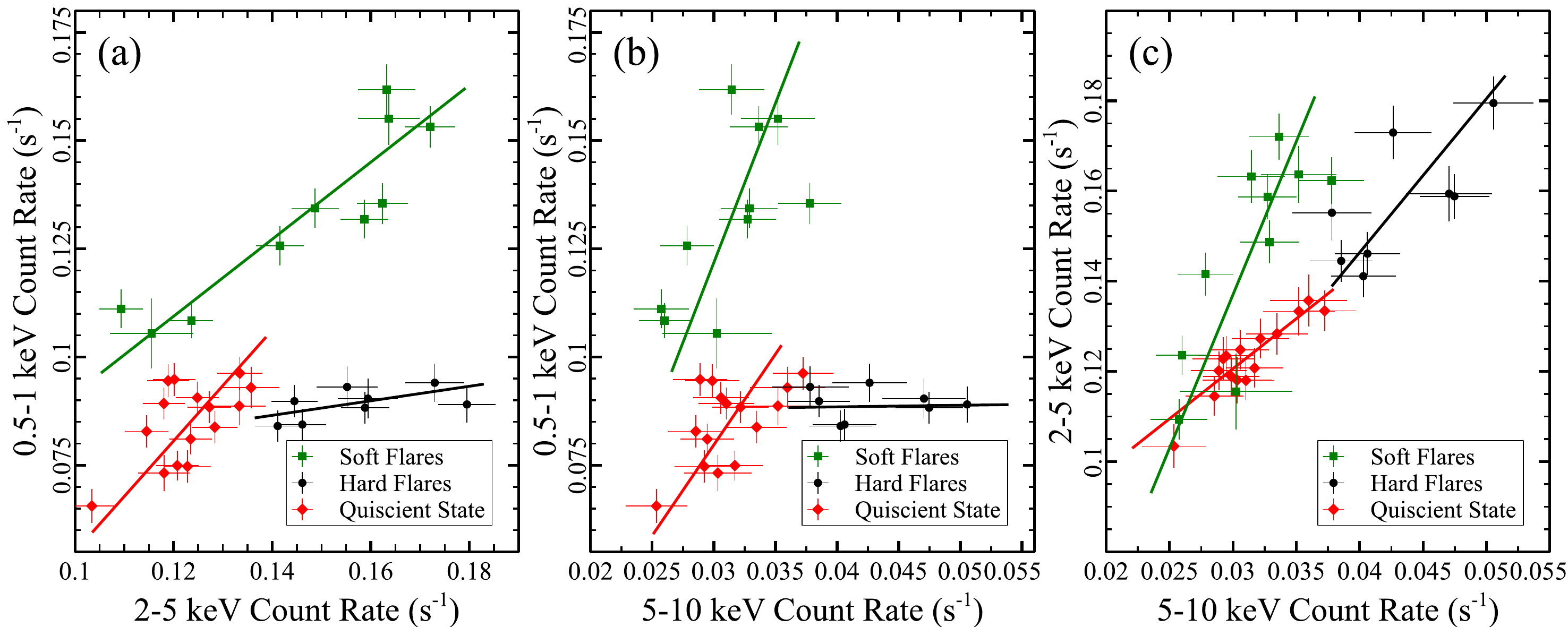}	
\caption{Flux-flux plots for \pds showing how the behaviour of the three segments are remarkably distinct. Panel (a): soft flux (0.5--1\,keV) against the mid flux (2--5\,keV). Panel (b): the soft flux against the hard flux (5--10\,keV). Panel (c): mid-flux flux against the hard flux. The three distinct grouped segments are plotted together in each panel showing the hard flares (black), quiescent (red) and the soft flares (green) where the corresponding best fit linear regression lines obtained separately are shown as solid lines respectively. The overall binning of the data is 2$\times$ orbital time (i.e., $11520$\,s) in order to improve the S/N ratio. The results of the linear correlations are listed in Table\,\ref{tab:gradients}.}    
\label{fig:pds_2007_count_count}
\end{figure}

In order to examine further how the different energy bands interact with one another, we specifically investigated the correlation between the flux (in terms of count rate) in the soft and the hard bands. On the basis of what we observed in the light curves and NSRs, we initially investigated the flux-flux distribution between the 0.5--1\,keV and 5--10\,keV bands (plotted in Fig.\,\ref{fig:pds_2007_count_count_6slice}) where the corresponding spectral states are clearly labelled. Within this distribution, we note that the soft flares (E, F), hard flares (A, C) and quiescent periods (B, D), lie in three distinct areas of the plots, with substantial overlap between each of the companion slices, e.g. between the soft flares E and F slices. Therefore given the similar behaviour between the hard flares, soft flares and quiescent periods, the A + C, E + F and B + D segments were separately combined to form single hard flare, soft flare and quiescent spectra respectively. This additional grouping was also done in order to achieve a higher signal-to-noise ratio (S/N) and they are adopted in the subsequent time-sliced spectral fitting in Section\,\ref{subsec:Time Dependent Spectral Analysis}.
\\
\indent In Fig.\,\ref{fig:pds_2007_count_count} we show three flux-flux plots corresponding to: 0.5--1\,keV (soft band) versus 2--5\,keV (mid-flux) in panel\,(a), 0.5--1\,keV versus 5--10\,keV (hard band) in panel\,(b) and 2--5\,keV versus 5--10\,keV in panel\,(c). Here as above the respective data were grouped into the hard flare (black), soft flare (green) and quiescent periods (red). Furthermore in order to increase the S/N, the data were binned to two-orbital size (11520 s). Subsequently, we computed the linear regression fits (in the form of $y=mx+C$) to the respective segments performed through the bivariate correlated errors and intrinsic scatter algorithm (BCES, \citealt{AkritasBershady96})\footnote{This is one of the most common methods that takes into account errors in both x and y values.} and tabulated in Table\,\ref{tab:gradients}. 
What emerges from all three flux-flux plots is that the three individual segments behave rather distinctively when compared with one another. In particular, Fig.\,\ref{fig:pds_2007_count_count} (a) and (b) emphasize how the flux-flux distribution in the hard flares largely follows an almost constant relation with a very small gradient (see Table\,\ref{tab:gradients} for values), with little rise in the soft flux compared to the mid and hard flux bands, strongly suggesting that the increase in flux is completely dominated by the hard band. On the other hand, the soft flares distribution shows the opposite behaviour, with a very steep relation between the 0.5--1\,keV vs 5--10\,keV flux. This reaffirms the behaviour seen in the light curves, where there is a considerably greater increase in soft flux compared to the hard band in the last part of the \suzaku observation at $t>250$\,ks. The behaviour of the quiescent period is intermediate between the soft and hard flares.

The behaviour of the three segments in the flux-flux plot between the 2--5 and 5--10\,keV energy band is not immediately apparent. Fig.\,\ref{fig:pds_2007_count_count}\,(c) reveals that the hard flares distribution appears to be an extension of the quiescent, as they are characterized by a comparable gradient ($m_{\rm hard\,flare}=3.4\pm0.9$ and $m_{\rm quiescent\,state}=2.2\pm0.4$), whereas the soft portion distribution is again steeper ($m_{\rm soft\,flare}=6.8\pm1.6$). Furthermore, these energy ranges provide more of a description of the behaviour of the power-law photon index, where a steeper $\Gamma$ would correspond to a higher 2--5/5--10\,keV ratio. The steepness of the soft flare portion of the flux-flux plot compared to the other two, is consistent with an increase in photon index corresponding to the soft flares. The origin of the contrasting spectral behaviour of the soft, hard and quiescent segments will be discussed in detail in the next section.

\begin{table*}

\begin{tabular}{c|ccccc}

\hline
                              \multirow{3}{*}{Panel}&    \multirow{3}{*}{$(x,y)$}                   &\textsc{bces} Linear regression          &\multicolumn{3}{c}{Grouped segments}\\

\\                              
                                   &                          &$y=mx+C$             &Soft Flares                 &Quiescent        &Hard Flares\\

\hline

\multirow{2}{*}{(a)}      &\multirow{2}{*}{2--5\,keV\,vs\,0.5--1\,keV}

                                                     &$m$                                &$0.89\pm0.14$               &$1.28\pm0.32$                   &$0.168\pm0.096$\\

										            &&$C$							      &$0.0026\pm0.0019$           &$-0.073\pm0.039$                &$0.063\pm0.015$\\

\\
\multirow{2}{*}{(b)}      &\multirow{2}{*}{5--10\,keV\,vs\,0.5--1\,keV}

                                                   &$m$                                 &$7.36\pm2.59$                  &$4.19\pm1.50$                 &$0.037\pm0.014$\\

										          &&$C$								 &$-0.099\pm0.081$               &$-0.046\pm0.045$              &$0.087\pm0.023$\\

\\

\multirow{2}{*}{(c)}      &\multirow{2}{*}{5--10\,keV\,vs\,2--5\,keV}

                                                   &$m$                                &$6.80\pm1.63$                   &$2.22\pm0.38$                &$3.41\pm0.91$\\

										          &&$C$							    &$-0.067\pm0.051$                &$0.054\pm0.012$              &$0.010\pm0.038$\\

\hline
\end{tabular}

\caption{Gradients and intercepts evaluated from the BCES linear regression fits corresponding to each segment plotted in Fig.\,\ref{fig:pds_2007_count_count} (a)--(c). From these values it is evident that the three segments behave distinctively. See text for more details.}
\label{tab:gradients}
\end{table*}
\pagebreak

\section{Time Dependent Spectral Analysis}
\label{subsec:Time Dependent Spectral Analysis}

In Section\,\ref{subsec:Modelling The Broadband SED} we used the \texttt{optxagnf} model to account for the broadband optical/UV to hard X-ray SED of \pds. In order to describe the lower S/N time-sliced 2007 \suzaku spectra over the 0.6--10\,keV band, we adopted a simpler two component model, in the form of a low energy blackbody plus a power law, to provide a more convenient parametrization of the intrinsic continuum. Although not physically motivated, the blackbody component here is a proxy for the cooler Comptonized emission responsible for the soft excess, as described by the \texttt{optxagnf} model. In particular, we want to test whether the broadband spectral variability is mainly produced by either (i) rapidly varying partial covering absorption, while the intrinsic continuum parameters are assumed to vary together in normalization, through the same scale factor, throughout the segments, or (ii) variations in the intrinsic shape of the continuum, such as the power law and soft excess.

\subsection{Partial Covering Changes}
\label{subsubsec:MODEL II - Neutral Partial Covering}

\begin{table*}
\begin{tabular}{cc|ccc|}

\hline

&                                           &Slice A + C                 &Slice B + D           &Slice E + F\\
\hline

\multirow{2}{*}{power law}

&$\Gamma$                                   &$2.45_{-0.04}^{+0.05}$      &$2.45^{\rm t}$        &$2.45^{\rm t}$\\

&norm$_{\rm po}^{\rm a}$                    &$4.14_{-0.38}^{+0.42}$      &$2.85\pm0.16$         &$2.87_{-0.29}^{+0.23}$\\

\\

\multirow{2}{*}{Blackbody}

&kT (eV)                                    &$100^{*}$                   &$100^{\rm t}$          &$100^{\rm t}$\\

&norm$_{\rm bb}^{\rm b}$                    &$8.42$                      &$5.80$                 &$5.85$\\

\\

\multirow{2}{*}{\texttt{pc}$_{\rm low}$} 

&$\log$($N_{\rm{H,low}}$/cm$^{-2}$)         &$21.5\pm0.2$                &$21.5^{\rm t}$         &$21.5^{\rm t}$\\

&f$_{\rm cov,low}$                          &$0.57_{-0.09}^{+0.18}$      &$0.37_{-0.07}^{+0.13}$ &$<0.12$\\

\\

\multirow{2}{*}{\texttt{pc}$_{\rm high}$}

&$\log$($N_{\rm{H,high}}$/cm$^{-2}$)        &$23.2_{-0.1}^{+0.3}$        &$23.2^{\rm t}$          &$23.2^{\rm t}$\\

&f$_{\rm cov, high}$                        &$0.34\pm0.05$               &$0.22\pm0.06$          &$<0.17$\\

\\
\hline

&Flux$_{0.5-2}^{\rm c}$                    &$3.09$                       &$2.84$                &$4.18$\\

\\

&Flux$_{2-10}^{\rm d}$                      &$4.19$                      &$3.16$                &$3.56$\\

\\

\hline

\multirow{2}{*}{Model Statistic}

&$(\chi^{2}/\nu)^{\rm e}$                   &$334/349$                   &$514/485$              &$471/449$\\

\\

&N.P.$^{\rm f}$                             &$0.71$                      &$0.17$                 &$0.22$\\

\\
&~~~~~~~~~~~~~Best Fit Statistic&$\chi^2/\nu=1320/1313$\\
\hline

\end{tabular}

\caption{\textit{Partial covering changes} parameters for the three \suzaku XIS 2007 combined segments. Here the spectral changes are accounted by the variability of the partial coverer covering fractions. The power-law and blackbody are varying together by the same scale factor throughout the segments. $^{\rm t}$ denotes that the parameter is tied during fitting, $^*$ indicates a parameter fixed during fitting.}
\vspace*{-5mm}
\begin{threeparttable}
\begin{tablenotes} 
	\item[a] Power-law normalization, in units of $10^{-3}$ ph keV$^{-1}$ cm$^{-2}$ s$^{-1}$,
	\item[b] blackbody normalization in units of $10^{-5}$ $(L_{39}/D^{2}_{10})$, where $L_{39}$ is source luminosity in units of $10^{39}$ erg s$^{-1}$ and D$_{10}$ is the distance to the source in units of 10 kpc,
	\item[c] overall absorbed flux, between 0.5--2\,keV, in units of $10^{-12}$ erg cm$^{-2}$ s$^{-1}$, 
    \item[d] overall absorbed flux between 2--10\,keV in units of $10^{-12}$ erg cm$^{-2}$ s$^{-1}$,	
	\item[e] $\chi^{2}$ and degrees of freedom calculated in each individual slice.
	\item[f] null hypothesis probability (N.P.) calculated in each individual slice.
\label{tab:modelII_3slice} 
\end{tablenotes}
\end{threeparttable}
\end{table*}

\begin{figure}
\centering	
\includegraphics[scale=0.70]{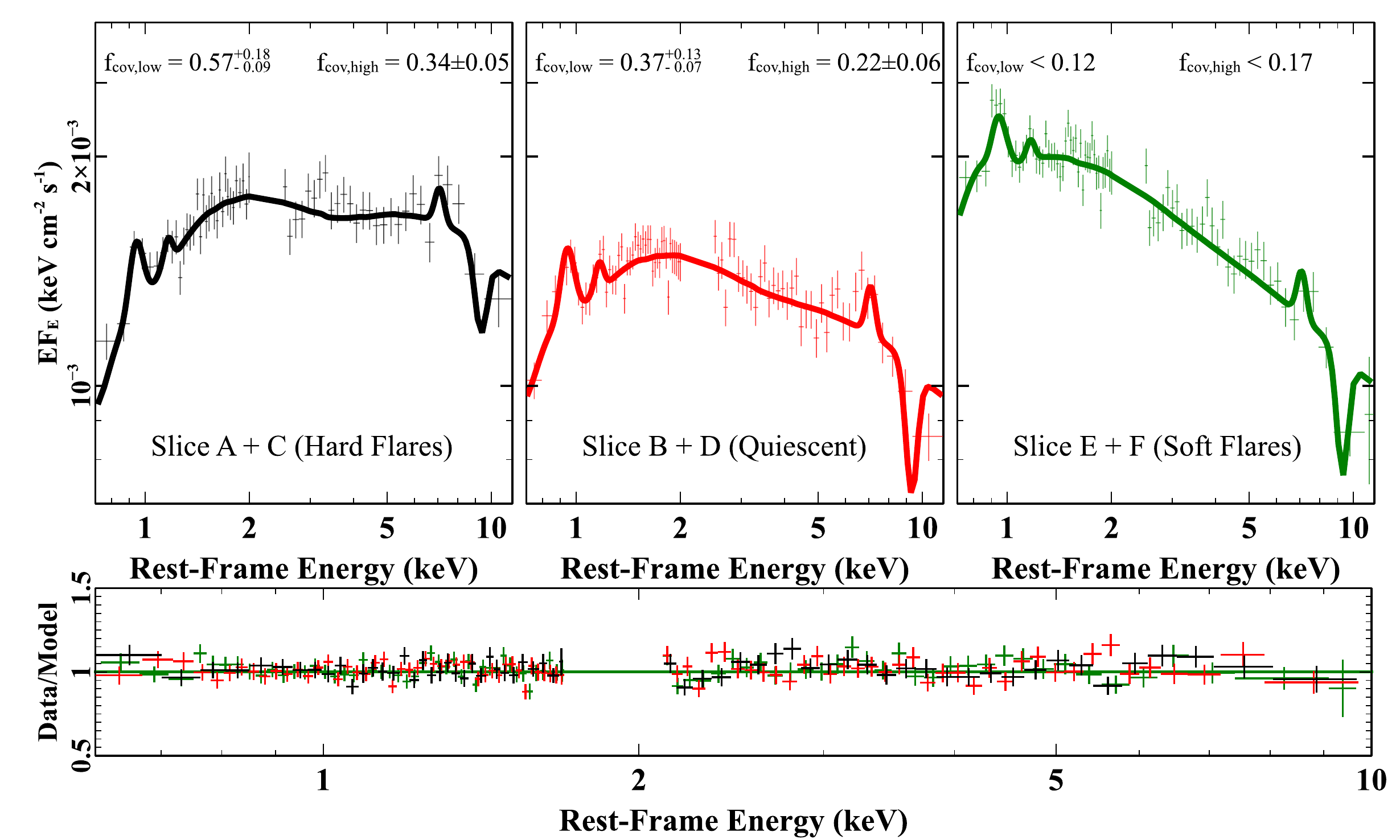}	
\caption{Plots of the XIS03 spectra from the \suzaku 2007 observation, showing the evolution from the hard flare (A + C) to the quiescent (B + D) and to the soft flare (E + F). Top panels: unfolded spectra for each of the three segments and their corresponding best-fit model corresponding to \textit{partial covering changes} model. In this scenario the spectral variability is caused by changes in the covering fractions of two absorbing zones ($f_{\rm cov,low}$ and $f_{\rm cov,high}$), where the rapid softening of the spectrum of slice E + F is caused by an uncovering of the X-ray emitting region. Bottom panel: the data/model residuals compared to the continuum for all three segments.}
\label{fig:pds456_2007_mo_ra_zpc}
\end{figure}

In the first scenario, the presence of compact clouds of gas reprocess the X-ray photons by partially absorbing the AGN emission, allowing a fraction ($1-f_{\rm{cov}}$) to emerge unattenuated. Previous studies of other AGN have found that these clouds can have a typical size-scale comparable to the X-ray emitting region, i.e. of the order of few tens of $R_{\rm g}$ \citep[e.g.,][]{Risaliti07}. The motivation behind this model is testing whether the observed change in flux (see Figs.\,\ref{fig:pds_2007_count_count_6slice} and \ref{fig:pds_2007_count_count}) that defines the three main spectral portions is the direct result of the presence of variable absorbers crossing the line-of-sight. The partial covering variability successfully explained the changes observed in the \suzaku 2013 campaign of \pds, when the source spectrum was much harder, although substantial intrinsic variability was also present (see M16 Sections 5.2.1 and 5.4). One interesting possibility is whether the harder portions of the 2007 observations could be explained by an increase in the absorber's covering fraction. Thus we constructed a \textit{partial covering changes} model of the form:

\begin{equation}
\begin{split}
\texttt{Tbabs}\times[\texttt{zpcfabs}_{\rm low}\times\texttt{zpcfabs}_{\rm high}\times(\texttt{po}+\texttt{bbody})+\texttt{zgauss}_{\rm em,soft}\\+\texttt{zgauss}_{\rm em,FeK}  +\texttt{zgauss}_{\rm abs,FeK}]  
\end{split}
\end{equation}

\noindent where \texttt{Tbabs} \citep{Wilms00} accounts for the Galactic absorption column density $N_{\rm H}=2\times10^{21}$ cm$^{-2}$. We apply this model to the three combined segments (A + C, B + D and E + F). For completeness, we adopted a Gaussian component (\texttt{zgauss}$_{\rm abs,FeK}$) to parametrize a single (due to the lower S/N present in the segments) iron\,K absorption profile with centroid rest-frame energy at $E=9.26_{-0.14}^{+0.16}$\,keV and with equivalent width $EW=-179_{-70}^{+56}$\,eV in A + C, $EW=-255_{-100}^{+80}$\,eV in B + D and $EW=-246_{-78}^{+96}$\,eV in E + F. The iron\,K absorption feature is almost constant throughout the three segments, which is in contrast to what was observed in the 2013 \suzaku spectra by M16, where the equivalent width of the iron\,K absorption line increased by a factor of $\sim10$ through the observation. As before we also included a Gaussian component (\texttt{zgauss}$_{\rm em,FeK}$) to model the iron\,K emission observed at slightly blueshifted energy of $E=7.08_{-0.16}^{+0.18}$\,keV with $EW=52_{-27}^{+30}$\,eV in A + C, $EW=73_{-37}^{+43}$\,eV in B + D and $EW=69_{-35}^{+40}$\,eV in E + F. We adopted a common velocity broadening of $\sigma=202_{-64}^{+92}$\,eV at the energy of the Fe\,K emission line, or $\sigma=265$\,eV at the energy of the Fe\,K absorption profile. From the time-averaged spectrum, we retained the two soft emission lines at $E=0.94_{-0.2}^{+0.1}$\,keV and $E=1.17\pm0.2$\,keV that are parametrized with two Gaussian components (\texttt{zgauss${_{\rm em,soft}}$}), with equivalent width and velocity broadening values consistent to the earlier broadband SED fit (see Table\,\ref{tab:sed}). 
\\
\indent When modelling the broadband SED in Section\,\ref{subsec:Modelling The Broadband SED}, the time-average 2007 spectrum required only a single thin layer of partial covering (see Table\,\ref{tab:sed}). However, when fitting the three segments the model prefers an additional layer improving the fit by $\Delta\chi^{2}/\Delta\nu=70/5$ ($\sim7.4\sigma$ confidence level), where the corresponding $F$-test chance improvement probability (compared to the model with only one partial coverer) is very low, with $P_{\rm f}=5.2\times10^{-12}$. This additional layer is required in this case to account for the hard to soft spectral changes. For simplicity, the $N_{\rm H}$ of the two partial covering zones are not allowed to vary between the three segments, so that the spectral changes over the course of the observation are only due to variations in the covering fractions\footnote{Note that we would obtain statistically equivalent results by letting the column densities, rather then covering fractions, vary between the segments.}. Similar to the \suzaku 2013 observation, the partial covering absorbers consist of a low (\texttt{zpcfabs}$_{\rm low}$) and high (\texttt{zpcfabs}$_{\rm high}$) column zone with column densities of $\log(N_{\rm H,low}/{\rm cm^{-2}})=21.5\pm0.2$ and $\log(N_{\rm H,high}/{\rm cm^{-2}})=23.2_{-0.1}^{+0.3}$ respectively. In this scenario we assume that there is no intrinsic continuum spectral variability, thus the relative flux normalizations of the blackbody and the power-law continuum were allowed to vary together by the same scale factor. The temperature of the blackbody component was fixed at $kT=100\,\rm eV$, however a value of $kT=94_{-22}^{+18}$\,eV was found when the temperature was allowed to vary.
\\
\indent For the low column zone, the covering fraction is substantially variable between the segments, reaching its highest value at $f_{cov,\rm low}=0.57_{-0.09}^{+0.18}$, while its minimum value was reached during the soft flares at $f_{cov,\rm low}<0.12$ (See Table\,\ref{tab:modelII_3slice}). For the high column the covering fraction ranges from its maximum value during the hard flares with $f_{cov,\rm high}=0.34\pm0.05$ to its minimum value during the soft flares with $f_{cov,\rm high}<0.17$. The high column zone is less variable than the low column zone, but the behaviour is similar. Nonetheless in order to compensate for the higher absorption, the intrinsic normalization of the continuum is required to be $\sim50\%$ higher during the hard flares, compared to either the soft and quiescent periods (see Table\,\ref{tab:modelII_3slice}). Overall, the \textit{partial covering changes} model provided an excellent fit to the data, with $\chi_{\nu}^{2}=1320/1313$ (see Fig.\,\ref{fig:pds456_2007_mo_ra_zpc}).

\begin{figure}
\centering	
\includegraphics[scale=0.70]{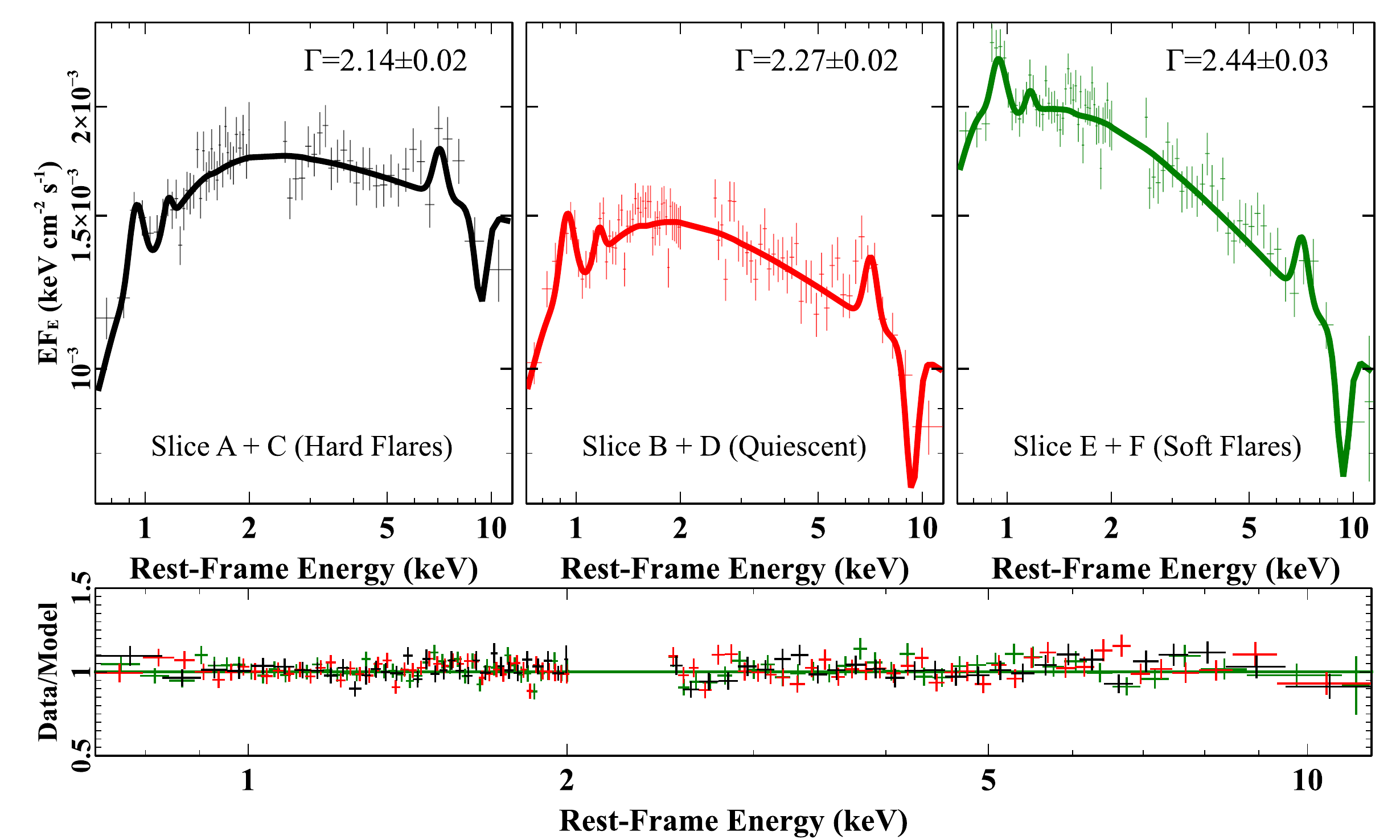}	
\caption{As per Fig.\,\ref{fig:pds456_2007_mo_ra_zpc}, but instead the spectral evolution compared to the \textit{intrinsic continuum changes} model. Note how the spectra soften during slice E + F which is mainly caused by an increase of the photon index together with the blackbody normalization (see Table\,\ref{tab:modelI_3slice}).}
\label{fig:pds456_2007_mo_ra}
\end{figure}

\subsection{Intrinsic Continuum Changes}
\label{subsubsec:MODEL I - Intrinsic Continuum Changes}

Alternatively, we test whether the observed short-term spectral variability is purely driven by intrinsic continuum changes. In this case, we assumed that the overall continuum is unabsorbed, with no modification by any partial covering absorbers. The parametrization of both the iron\,K absorption and emission profiles, together with the two soft emissions lines are perfectly consistent with the previous model. In order to account for the variability between the three segments, we parametrized the spectra with a baseline model of the form:

\begin{equation}
\begin{split}
\texttt{Tbabs}\times(\texttt{bbody}+\texttt{powerlaw}+\texttt{zgauss${_{\rm em,soft}}$}\\+\texttt{zgauss${_{\rm em,Fe\,K}}$}+\texttt{zgauss${_{\rm abs,Fe\,K}}$}).  
\end{split}
\end{equation}

\noindent Here, the power-law photon index ($\Gamma$) and the corresponding power-law and blackbody normalizations were allowed to vary independently between the three segments. The photon indices were indeed variable, yielding $\Gamma=2.14\pm0.02$, $\Gamma=2.27\pm0.02$ and $\Gamma=2.44\pm0.03$, for the hard flare, quiescent and soft flare portions respectively. The resulting model is plotted in Fig.\,\ref{fig:pds456_2007_mo_ra} and clearly shows that the power-law photon index generally increases as the observation progresses, being flatter during the hard segment and becoming steeper during the soft segment and intermediate in slope between the two in the quiescent segment. The spectral shape is therefore largely consistent to what was observed in the flux-flux plots in Fig.\,\ref{fig:pds_2007_count_count_6slice} and Fig.\,\ref{fig:pds_2007_count_count}, presented earlier in Section\,\ref{subsec:Flux-Flux plot}.  
\\
\indent As with the previous model, for consistency, the soft excess was also parametrized with a blackbody component where its temperature was fixed at $kT=100\,\rm eV$ (which if allowed to vary was found to be $kT=98\pm16$\,eV). In this fit, the blackbody normalization increased by a factor of $\sim3.5$ across the three segments (i.e., from A + C to E + F) as the overall observation increasingly became softer. Overall in this scenario (as well as in the partial covering changes case) we obtained a statistically excellent fit to the data i.e., $\chi^{2}/{\nu}=1328/1316$. This suggests that the spectral variability in this 2007 \suzaku observation of \pds can be statistically equally well explained by both scenarios.

\begin{table*}
\begin{tabular}{cc|ccc|}

\hline

&                             &Slice A + C               &Slice B + D         &Slice E + F\\
\hline

\multirow{2}{*}{power law}

&$\Gamma$                                &$2.14\pm0.02$              &$2.27\pm0.02$       &$2.44\pm0.03$\\

&norm$_{\rm po}^{\rm a}$                 &$2.06\pm0.05$              &$1.86\pm0.03$       &$2.60\pm0.05$\\

\\

\multirow{2}{*}{Blackbody}

&kT (eV)                                 &$100^{*}$                 &$100^{\rm t}$        &$100^{\rm t}$\\

&norm$_{\rm bb}^{\rm b}$                 &$<1.76$                   &$1.69\pm0.80$        &$5.92^{+1.14}_{-1.11}$\\

\\

\hline

&Flux$_{0.5-2}^{\rm c}$                    &$3.09$                       &$2.84$                &$4.18$\\

\\

&Flux$_{2-10}^{\rm d}$                      &$4.19$                      &$3.16$                &$3.56$\\

\\

\hline

\multirow{2}{*}{Model Statistic}

&$(\chi^{2}/\nu)^{\rm e}$             &$338/353$                     &$516/490$            &$474/454$\\

\\

&N.P.$^{\rm f}$                                   &$0.71$                        &$0.20$               &$0.25$\\

\\
&~~~~~~~~~~~~~Best Fit Statistic&$\chi^2/\nu=1328/1317$\\
\hline

\end{tabular}

\caption{\textit{Intrinsic continuum changes} parameters for \suzaku XIS 2007 three combined segments. Here the power-law and blackbody components are allowed to vary independently accounting for the short-term spectral variability. $^{\rm t}$ denotes that the parameter is tied during fitting, $^*$ indicates a parameter fixed during fitting}
\vspace*{-5mm}
\begin{threeparttable}
\begin{tablenotes} 
	\item[a] Power-law normalization, in units of $10^{-3}$ ph keV$^{-1}$ cm$^{-2}$ s$^{-1}$,
	\item[b] blackbody normalization in units of $10^{-5}$ $(L_{39}/D^{2}_{10})$, where $L_{39}$ is source luminosity in units of $10^{39}$ erg s$^{-1}$ and D$_{10}$ is the distance to the source in units of 10 kpc,
	\item[c] overall absorbed flux, between 0.5--2\,keV, in units of $10^{-12}$ erg cm$^{-2}$ s$^{-1}$, 
    \item[d] overall absorbed flux between 2--10\,keV in units of $10^{-12}$ erg cm$^{-2}$ s$^{-1}$,	
	\item[e] $\chi^{2}$ and degrees of freedom calculated in each individual slice.
	\item[f] null hypothesis probability (N.P.) calculated in each individual slice.
\label{tab:modelI_3slice} 
\end{tablenotes}
\end{threeparttable}

\end{table*}

\begin{figure}
  \centering
\includegraphics[width=0.5\textwidth]{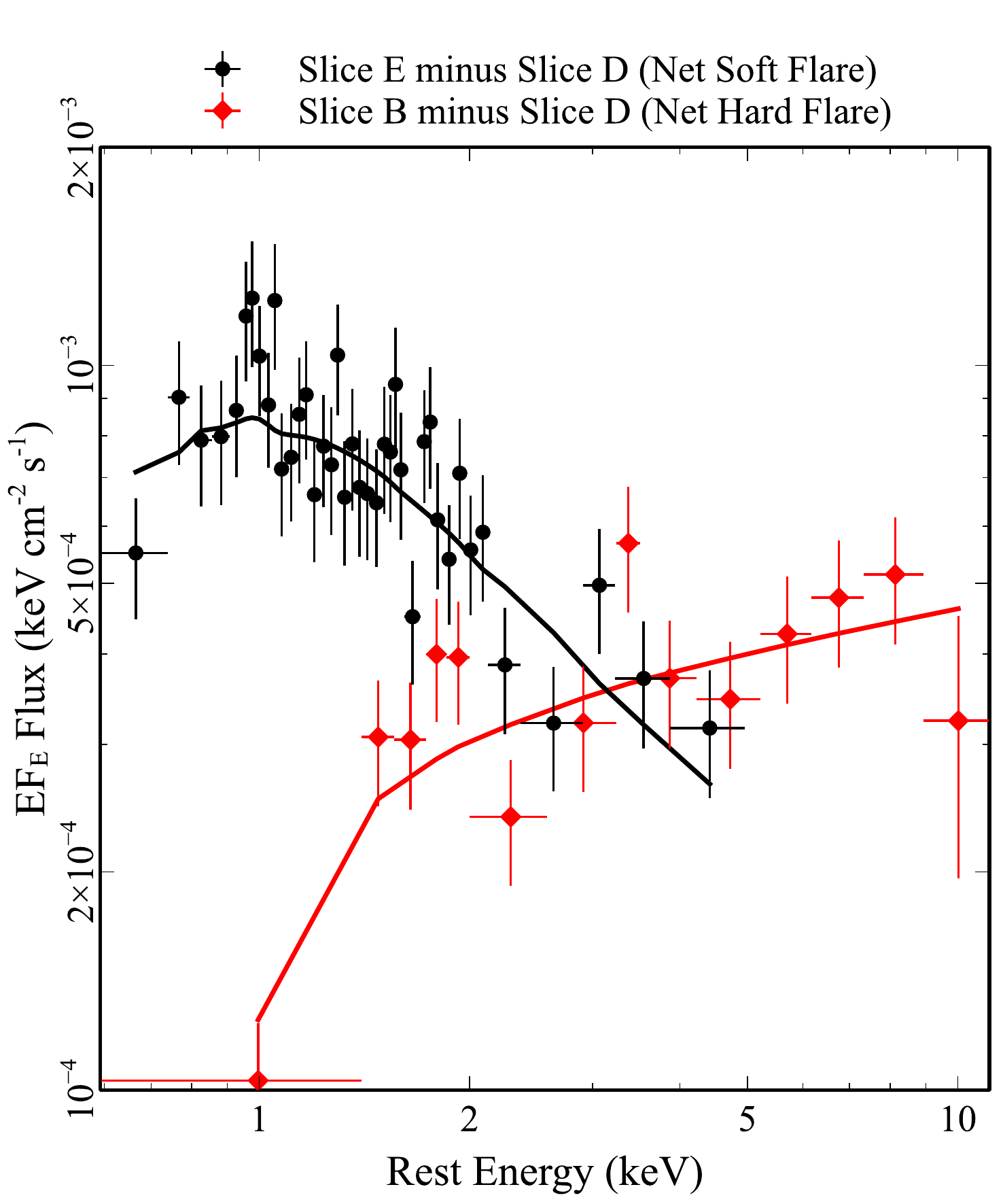}
  \caption{The net soft flare (E -- D, black) and the net hard flare (C -- D, red) difference spectra. The solid black and red lines correspond to the dual power law model which in the net soft spectrum is characterized by a very soft photon index $\Gamma\sim3.8$ with a weak `harder' tail at $\Gamma\sim2.3$; while the net hard spectrum is characterized by a very hard photon index of $\Gamma\sim1.8$. It is evident that the observation is characterized by the variability of two distinct continuum components. Both net spectra were grouped to a 5$\sigma$ detection per bin.}
  \label{fig:pds_2007_flare_soft_min_qstate}
\end{figure}

\subsubsection{Difference Spectrum Analysis}
\label{net_flares}

We have established that the spectral variations during the 2007 \suzaku observation are characterized by two prominent hard and soft events. In order to characterize the properties of the variability in the different soft versus hard segments, we calculated the corresponding difference spectra. This was accomplished by simply subtracting (in turn) the individual slice D spectrum (the quiescent portion of the observation lowest in flux) from the adjacent slice C (hard flare) and likewise from slice E (soft flare). Thus these subtracted spectra are now referred to as net hard flare (C -- D) and net soft flare (E -- D). We found that the net hard flare spectrum could be fitted with a simple power-law model component with a relatively hard photon index (in the context of \pds) of $\Gamma=1.8\pm0.2$. The net soft flare spectrum required a power-law component consisting of a very steep photon index at $\Gamma=3.8\pm0.5$ dominating below 2\,keV, together with a weak `harder' tail\footnote{The presence of a weak `harder' tail is consistent with the behaviour of the 2--5\,keV band light curve in Fig.\,\ref{fig:pds_2007_3lc_hr}\,(P.3) where the two prominent soft flares in the second half of the observation, that dominates the 0.5--1\,keV band, are also still detected in the 2--5\,keV band.} ($\Gamma\sim2.3$) mainly affecting the 2--5\,keV band. Both cases produce an acceptable fit with $\chi^{2}/\nu=51.6/48$ and the resulting difference spectra with their corresponding power-law components are plotted in Fig.\,\ref{fig:pds_2007_flare_soft_min_qstate}.
\\
\indent We find that both the net soft and net hard flare spectra could have been equally fitted with a single \texttt{compTT} model \citep{Titarchuk94}, each characterized by a distinct Comptonizing region. The net soft flare portion can be reproduced by an optically thick (with $\tau=2.0\pm0.2$) `warm' Comptonizing region with temperature of $kT=4.7_{-0.4}^{+0.5}$\,keV. On the other hand, an optically thin ($\tau=0.34_{-0.04}^{+0.05}$) `hot' Comptonizing region with temperature fixed at $kT\sim100$\,keV, characterizes the net hard flare. This produces an overall acceptable fit with $\chi^{2}/\nu=55.7/50$. This result suggests that the overall spectral variability of \pds in 2007 is likely to be characterized by the superposition of two different variable continuum components; (i) variable soft flares (or `warm' coronal component) and (ii) variable hard flares (or `hot' coronal component).

\subsection{Fractional Variability}
\label{subsec:Fractional Variability}

\begin{figure}
  \centering
\includegraphics[width=1.0\textwidth]{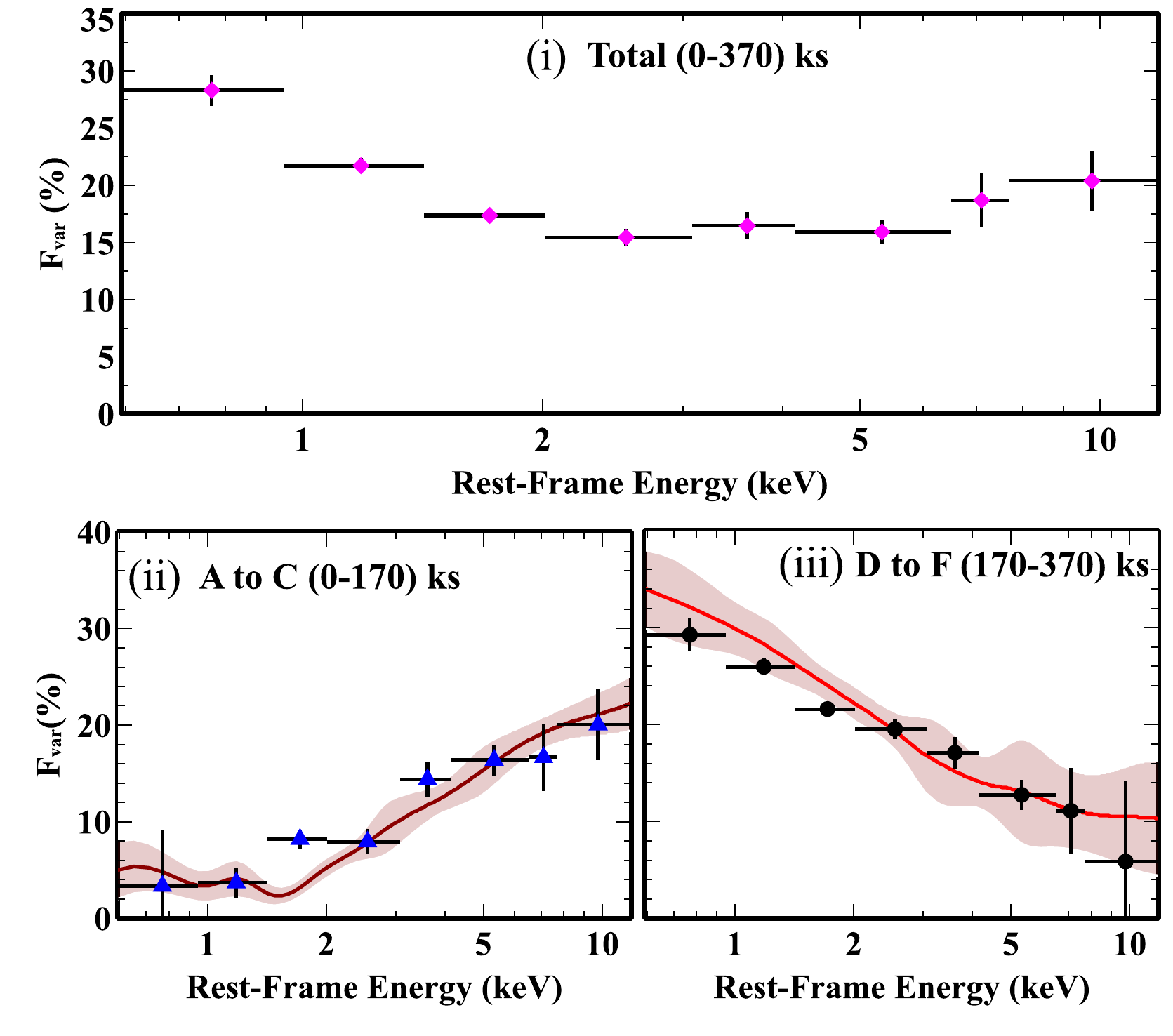}
\caption{Fractional X-ray variability ($F_{\rm var}$) from the 2007 \suzaku observation where panel (i) shows the total $F_{\rm var}$ spectrum, panel (ii) shows the $F_{\rm var}$ computed only in the first half of the observation while panel (iii) the $F_{\rm var}$ from only the second half of the observation. It is clear that the total $F_{\rm var}$ (i) is the superposition of (ii) and (iii) strongly suggesting that the observation is dominated by distinct variable hard and soft components. The solid red lines correspond to the simulated $F_{\rm var}$ shape based on the \textit{intrinsic continuum changes} model (see text for details). The curves have been smoothed with a spline function, and the shaded area indicates the $1\sigma$ dispersion for $3000$ mock light curves.} 
	\label{fig:fvar_2007_new}
\end{figure}

To further quantify the spectral changes, we computed the fractional variability ($F_{\rm var}$ hereafter) in different energy bands, adopting a time binning of $5760$\,s (i.e. one \suzaku orbit) and using the method described in \citet{Vaughan03}. The $F_{\rm var}$ spectrum computed over the entire observation is plotted in the top panel of Fig.\,\ref{fig:fvar_2007_new}\,(i). However a striking difference in the $F_{\rm var}$ spectrum was found when comparing the two halves of the 2007 observation.
\\
\indent The $F_{\rm var}$ spectra computed over the first (0--170\,ks between A to C) and the second part (170--370\,ks between D to F) of the observation, are plotted in Fig.\,\ref{fig:fvar_2007_new}\,(ii) and Fig.\,\ref{fig:fvar_2007_new}\,(iii) respectively. Their resulting shapes were remarkably similar to the difference spectra plotted in Fig.\,\ref{fig:pds_2007_flare_soft_min_qstate}, where in panel (ii) the variability was most prevalent in the hard energy band (hard flare portion) and conversely in panel (iii), most of the variability occurred in the soft band (soft flare portion) of the observation. This confirms the two component behaviour represented by different soft and hard emission components. The resultant shape of the total $F_{\rm var}$ spectrum, in panel (i), is therefore consistent with the contribution of both components superimposed. One possibility is that the first half of the observation is dominated by changes in the intrinsic power-law flux and slope, resulting in enhanced variability above 2\,keV, while the behaviour of the second half of the observation is dominated by a drastic rise in the soft X-ray emission below 2\,keV.
\\
\indent In order to test more quantitatively what we have found above, simulations were used to test the possible variability mechanism that produced both $F_{\rm var}$ shapes in Fig.\,\ref{fig:fvar_2007_new}\,(ii) and (iii). This was achieved by computing two separate $F_{\rm var}$ spectra, each representing one half of the observation, from 3000 simulated light curves. More specifically, each light curve in the observation has 65 data points (see Fig.\,\ref{fig:pds_2007_3lc_hr}), which is effectively produced by having 65 separate spectra per each simulated light curve; in other words $3000*65=195000$ spectra were randomly generated within the input ranges extrapolated from the earlier best fit \textit{intrinsic continuum changes} model. We computed two sets of simulations by adopting for simplicity a simple dual power-law component, described earlier in Section\,\ref{net_flares}, as the input model. For the first half of the observation (simulation A hereafter) only the hard power-law photon index, corresponding to the hard variable component, was allowed to vary within the range from $2.0\leq\Gamma\leq2.5$. This is consistent with the observed $\Gamma$ changes in Section\,\ref{subsubsec:MODEL I - Intrinsic Continuum Changes}, with no change required in the soft component. For the second half of the observation (simulation B hereafter), we only allowed the normalization of the soft power-law component (of photon index $\Gamma=4.0$) to vary between $1$--$9\times10^{-4}$\,ph\,keV$^{-1}$\,cm$^{-2}$\,s$^{-1}$. This can be seen in Fig.\,\ref{fig:pds_2007_flare_soft_min_qstate} from the respective photon fluxes at 1\,keV in the difference spectra. 
\\
\indent The resulting simulations are overlaid on the $F_{\rm var}$ spectra in Fig.\,\ref{fig:fvar_2007_new}\,(ii) and (iii), where the red solid lines correspond to the simulated $F_{\rm var}$ model and associated $1\sigma$ dispersion, shown as the shaded area, compared to the actual $F_{\rm var}$ spectrum measured from the observations. We found that the observed shape of the $F_{\rm var}$ spectrum plotted in Fig.\,\ref{fig:fvar_2007_new}\,(ii) is well reproduced by simulation A with a moderate pivoting of the photon index. The rising shape of the $F_{\rm var}$ spectrum towards higher energies can be pictured as a variable photon index due to the hardening of the spectrum. On the other hand, simulation B largely reproduced the $F_{\rm var}$ shape of the second half of the observation, plotted in Fig.\,\ref{fig:fvar_2007_new}\,(iii), achieved by a variable normalization of the `soft' component. Therefore a change in normalization of the soft component in the 2007 D to F segments, would have been mainly responsible for an $F_{\rm var}$ spectrum dominated by variability in the soft band rather than at higher energies. The overall behaviour is consistent with a two component variable continuum where the soft is dominated by a rapid increase in the normalisation of the soft component, causing the overall steepening of the spectrum, whereas during the first part of the observation, when the soft component is less prevalent, the photon index of the power law tail tends towards harder values.  
\\
\indent On the other hand it may be difficult to reconcile the distinct behaviour in the $F_{\rm var}$ spectra in Fig.\,\ref{fig:fvar_2007_new}, from the two halves of the observation, with the variable \textit{partial covering changes} model. Although we do not test this explicitly here, the only way for this bimodal behaviour to be reproduced might be for the first half of the observation to be dominated only by the variability of the high column absorption zone, which then would primarily enhance the variability in the harder band. Then in the second half of the observation, the variability would need to be dominated only by the lower column partial coverer, enhancing the variability in the soft band. Nonetheless such a distinct behaviour of the two absorption zones may seem somewhat contrived and furthermore, as discussed in the next section, such rapid absorption variability is likely to be on too short timescales to be physically plausible. Finally, during 2013 \suzaku observations, when \pds was in low-flux state dominated by absorption, the $F_{\rm var}$ spectrum did not show this simple power-law like variability behaviour from either of two distinct continuum components (see M16 Fig.\,11); in those observations the $F_{\rm var}$ spectrum had pronounced curvature and showed enhanced variability in the iron\,K band due to the variability of the outflow.

\section{Discussion}
\label{sec:Discussion}

In the previous sections we found that the observed short-term spectral variability can be statistically well explained by either: (i) the presence of variable partial covering absorption, or (ii) an intrinsically variable continuum. In scenario (i), the X-ray spectral variability may be caused by an uncovering of the intrinsically steep continuum. This is due to changes in the covering factors from two zones of partial covering absorbers, characterized by the same physical properties observed in the 2013 \suzaku campaign (see M16). On the other hand in scenario (ii), such a rapid spectral variability may be caused by intrinsic flaring, with some episodes (A + C) intrinsically harder and others (E + F) intrinsically softer. The latter scenario can also reproduce the $F_{\rm var}$ spectra discussed above. In this section we will discuss the feasibility and physical implications of these two scenarios.

\subsection{Variable Partial Covering}
\label{subsec:Variable Partial Covering}

In Section\,\ref{subsubsec:MODEL II - Neutral Partial Covering} we described a scenario where the short-term spectral variability can be well explained by variable a partial covering model. However, there are some physical implications that need to be addressed. First of all, when modelling the broadband SED in Section\,\ref{subsec:Modelling The Broadband SED}, the time-averaged 2007 spectrum is not strongly absorbed as it allows only a weak single layer of partial covering with column density $\log(N_{\rm H}/{\rm cm^{-2}})=23.1_{-0.3}^{+0.1}$ and $\sim15\%$ covering fraction (see Table\,\ref{tab:sed}). Yet when fitting the three segments, two zones of partial covering are required in order to account for the observed spectral variability; where the low column is mainly affecting the soft energy band, whilst the high column is changing the spectral shape above 2\,keV. Furthermore despite being more absorbed, the hard flare segment is required to have an intrinsically higher continuum level (by a factor of $\sim50\%$, see Section\,\ref{subsubsec:MODEL II - Neutral Partial Covering}) compared to the less absorbed soft and quiescent periods. This may seem somewhat contrived, as the most absorbed portion of the observation is required to be at the highest flux level.
\\
\indent A physical issue that makes the partial covering scenario unfeasible is the variability timescale being too short. To determine whether the timescale of the partial covering changes are realistic, an estimate of the absorber velocity is required. \citet{Reeves16} presented a high resolution soft X-ray analysis of all \pds RGS spectra to date. By measuring the discrete features in the RGS spectra, they found the presence of X-ray absorbers moving at a typical outflow velocity of $v\sim0.20$--$0.25c$. Furthermore in M16, by fitting the 2013 broadband spectra of \pds, the high column partial coverer was also suggested to be outflowing at velocity of the order of $v_{\rm pc}\sim0.25c$. M16 argued that the partial coverer could be considered as a plausible, less ionized, component of the same fast wind. From adopting a transverse velocity (comparable to the outflow velocity i.e., $v_{\rm T}\sim v_{\rm pc}\sim0.20$--$0.25c$) for the partial covering absorber, we can determine the absorber size-scale. Assuming that an eclipsing event may be responsible for the observed flux drop between the two (fully unobscured) soft flares, in slice E and F (see Fig.\,\ref{fig:pds_2007_3lc_hr} P.2), the timescale at which this event occurs would be of the order of $\sim50\,\rm ks$, suggesting a typical size-scale of the eclipsing clump to be $\Delta R_{\rm clump}\sim v_{\rm pc}\Delta t\sim3$--$4\times10^{14}\,{\rm cm}\sim2$--$3\,R_{\rm g}$. Note this is also similar to the timescale of the `hard' events, where in the partial covering model the quasar becomes more obscured. Taking the column density of the variable absorber to be $N_{\rm H}\sim10^{22}$\,cm$^{-2}$, we obtain a value for the average hydrogen number density of $n_{\rm H}\sim N_{\rm H}/\Delta R\sim2\times10^{7}$\,cm$^{-3}$. To be conservative, we can also assume that these partial coverers are partially ionized, rather than neutral, with the ionization parameter required to be at most $\log(\xi/{\rm erg\,cm\,s^{-1}})\lesssim2.5$ (see M16) in order to remain opaque at soft X-rays~\footnote{Note that in section 5.4 in M16 the ionization parameter of the low and high column partially ionized partial covering absorbers was found to be $\log(\xi/{\rm erg\,cm\,s^{-1}})=0.62_{-0.08}^{+0.18}$ and $\log(\xi/{\rm erg\,cm\,s^{-1}})=2.5\pm0.2$.}. 
\\
\indent From the definition of the ionization parameter and taking the ionizing luminosity of \pds consistent with N15 and M16 to be $L_{\rm ion}\sim5\times10^{46}\,{\rm erg\,s^{-1}}$, we can estimate the radial location of these absorbers to be at $R=\left ( L_{\rm ion}/n_{\rm H}\xi \right )^\frac{1}{2}\gtrsim10^{18}\,{\rm cm}\gtrsim10^{4}\,R_{\rm g}$. This would place these clumps at parsec-scale distances from the black hole in \pds. In contrast \citet{Reeves16} found, in the RGS spectra, that these absorbers were located at least one order of magnitude closer (at BLR scale). Therefore the timescales at which we observe the uncovering of the X-ray source would lead to rather compact clumps (of a few $R_{\rm g}$ in extent) with typical size-scales physically too small to be located at such large (parsec-scale) distances. On the other hand, by assuming that the Keplerian velocity across the source is comparable to the outflow velocity i.e., $v_{\rm pc}\sim v_{\rm K}$, we can estimate the radial distance of these absorbers to be $r_{\rm clump}\sim\frac{c^2}{v_{\rm K}^2}\,R_{\rm g}\sim 20\,R_{\rm g}$. Thus if these clumps were located at such a small distance from the inner accretion disc, with a typical density of $n_{\rm H}\sim N_{\rm H}/\Delta R\sim2\times10^{7}$\,cm$^{-3}$, the ionization derived would be unrealistically high i.e., $\xi\sim L_{\rm ion}/n_{\rm H}{\rm R^2}\sim10^8\,{\rm \rm erg\,cm\,s^{-1}}$. Indeed this would be two orders of magnitude higher even than the high ionization absorber at iron\,K (N15). Thus given the compactness of the absorbing clouds and the high degree of ionization required to place them at a reasonable distance from the black hole, the partial covering scenario seems less plausible to explain the rapid spectral changes in \pds. Thus the rapid variability in these high-flux observations are more likely to be intrinsic in origin. In contrast the prolonged ($\sim1\,\rm Ms$) low-flux periods observed in 2013 with \suzaku are likely to be due to enhanced absorption, as is also required to explain the hard spectral shape during those observations as well as the prominent increase in depth of the iron\,K absorption profile (see \citealt{Gofford14} and M16).

\subsection{Intrinsic Continuum Variability}

All the observational evidence in this work suggests there are two sources of variability present in this 2007 \suzaku observation of \pds in the form of hard and soft variable components. Note that this two-component variability has also been detected in other AGN by means of principal component analysis \citep[e.g.,][]{Miller07,Parker15}. In the \pds observation the spectrum is initially hard, becoming softer as the observation progresses. A possible explanation can be attributed to a large increase of the soft photon flux, detected below $\sim2$\,keV, towards the end of the observation, which may act as the `seed' photons that produce the hard power-law component. In particular an increase in soft photon flux could lead to the cooling of the `hot' coronal electrons, which are responsible for producing the hard X-ray power-law via Compton up-scattering. Thus an increase in soft photon flux may lead to a steepening of the power-law photon index (from $\Gamma\sim2$ to $\Gamma\sim2.5$), as seen in the later E + F segment of the observation.

\subsubsection{Compton Cooling of the Corona}
\label{subsec:Compton Cooling}

Here we investigate whether any Compton cooling of the corona could occur on a similar timescale to the spectral changes seen in the 2007 \suzaku observation. In Section\,\ref{subsubsec:MODEL I - Intrinsic Continuum Changes}, we determined that the softening of the spectrum occurs on a typical timescale of $\sim50$\,ks. In comparison, the cooling rate (i.e., the rate of energy loss) per electron from inverse Compton scattering (IC) can be expressed as:

\begin{equation}
\left (\frac{\mathrm{d} E}{\mathrm{d} t} \right )_{\rm IC}=\frac{4}{3}\sigma_{\rm T}c\left ( \frac{v_{\rm e}}{c} \right )^{2}\gamma^{2}U_{\rm rad}
\end{equation}

\noindent where the $\sigma_{\rm T}\sim6.65\times10^{-25}$ cm$^{2}$ is the Thomson cross section of an electron, $v_{\rm e}$ is its velocity, $\gamma$ is the Lorentz factor, and $U_{\rm rad}=\frac{L}{4\pi r^{2}c}$ is the energy density of the incident radiation. From recent studies performed on Type 1 AGN samples we can assume a typical coronal temperature of the order $\sim100$\,keV, based on their high energy roll over \citep[e.g.,][]{Derosa12,Ursini15}. It follows that a Lorentz factor of $\gamma\sim1.2$ is required in order to produce such kinetic energies, and thus $\left ( \frac{v_{\rm e}}{c} \right )^{2}\sim1/3$. 
\\
\indent The radiation field ($U_{\rm rad}$ hereafter) is estimated in terms of the intrinsic soft X-ray luminosity, in the 0.5--2\,keV band, assuming that the soft X-ray component acts as the input `seed' photons for the hard power-law. The soft X-ray luminosity extrapolated from soft flare segment is $L_{0.5-2}\sim10^{45}$ erg s$^{-1}$, which may be a factor of a few higher if the soft component extends below the 0.5\,keV low-energy bandpass. Thus, our choice of energy range provides a conservative estimate of $U_{\rm rad}$. From the observed timescale of the variability of $\Delta t\sim50$\,ks, we can deduce the typical size of the X-ray emitting region to be $r\sim10\,R_{\rm g}\sim1.5\times10^{15}$ cm in \pds. This yields an energy density of the radiation field in the order of $U_{\rm rad}\sim1000$ erg cm$^{-3}$. Thus we are now able to estimate the cooling rate, of a single electron, due to inverse Compton scattering to be:

\begin{equation}
\left ( \frac{\mathrm{d} E}{\mathrm{d} t} \right )_{\rm IC}\sim1.3\times10^{-11}\,{\rm erg\,s^{-1}}.
\end{equation}

As noted above, the most popular models predict that the high-energy cut-off ($E_{\rm cut}$) of the electron population in the corona is of the order of $kT_{\rm e}\sim100$\,keV \citep[e.g.,][]{Marinucci14}. On this basis we are able to estimate the corona Compton cooling time, i.e.,  $t_{\rm cool}=\dfrac{E_{\rm kin}}{\left ( \frac{\mathrm{d} E}{\mathrm{d} t} \right )_{\rm IC}}$, where $E_{\rm kin}\sim2$--$3\,kT_{\rm e}$ is the typical energy of the coronal electrons. Thus we estimate that the cooling time to be of the order of $t_{\rm cool}\sim40\,{\rm ks}$. Indeed this is on a similar timescale to where we observe the spectrum (segment E + F) becoming softer during the two flares. Thus the increase in photon index of the primary power-law continuum may occur as a result of the strong soft X-ray flares in the latter half of the 2007 observation.

\subsubsection{Energetics and Reprocessing}
\label{Energetics and Reprocessing}

Although it is not possible to determine the coronal geometry, one possibility is that it is characterized by a thick `warm' atmosphere blanketing the inner region of the disc, responsible for the soft excess and by a compact, but thin, `hot' coronal region located above the accretion disc which is responsible for the high-energy power-law \citep{Done12}. However, due to the rapid soft band variability observed here, the soft (as well as the hard) variable component cannot be overly extended (see Section\,\ref{subsec:Properties Of The Flares}). 
\\
\indent What might occur in this `dual-layered' configuration, is a build up of energy stored in the `hot' corona, previously estimated to be in the order of $\sim10^{51}\,{\rm erg}$ over timescales of $\sim10^{5}$\,s from the duty cycle of the X-ray flares \citep{Reeves02}. This is followed by some `triggering' factors, perhaps in the form of `cascade of  events' \citep{Merloni01,Reeves02}, which would eventually dissipate the energy stored, leading to the hard X-ray flares as seen during segment A + C. Note that similar X-ray hard flares were also found in earlier \textit{RXTE} and \textit{BeppoSAX} observations of \pds\citep{Reeves00,Reeves02}. In turn these hard photons may illuminate the outer layers of the disc, leading to an increase in temperature of the `warm' scattering region. By doing so, the `heated' electrons would, effectively, impart more energy to the EUV disc-photons undergoing Compton up-scattering into the observable soft X-ray domain as seen during the soft flares (segment E + F). Therefore a relatively small increase in temperature can in principle lead to a large increase in the soft X-ray flux as the Comptonized tail of the disc is shifted into the observable band above $\sim0.5$\,keV.
\\
\indent This may be regarded as an over simplification of a more complex physical scenario; nonetheless the underlying idea is that the hard events are responsible for the energy injection into the corona, whereas the soft events contribute to its energy loss as an increasing flux of soft photons is supplied into it. Subsequently, the hot electrons give up their energies via inverse Compton scattering to the soft photons. Eventually  what we may expect is that further hard events would re-inject energy into the system (i.e., coronal heating) resulting in the hardening of the spectra defining a possible onset of a new cycle.

\section{conclusion}

In this paper, we have presented the results from a \suzaku observation carried out in 2007 (total duration $\sim370\,{\rm ks}$) when \pds was in an unabsorbed state. We have detected the presence of strong hard and soft band X-ray flaring on timescales of $\sim50\,\rm ks$, confined in the first and second half of the observation respectively. We investigated the time-averaged broadband continuum by constructing a spectral energy distribution (SED), where we included the optical photometric data (OM) and hard X-ray data from observations of a similar spectral state carried out in 2013 with \xmm \& \nustar. We have found that the self-consistent (accretion disc and corona) \texttt{optxagnf} model was able to explain successfully the SED with values consistent with what was found in the later \suzaku 2013 campaign of \pds. However, in this 2007 \suzaku observation we appear to observe the bare continuum from \pds, with little intrinsic absorption.
\\
\indent The short-term spectral variability could be interpreted in terms of either: (i) partial covering changes or (ii) intrinsic continuum variability. We have found that statistically speaking both scenarios produce equivalent fits to the data. However the partial covering scenario implies that such short timescale variability ($\Delta t\sim50$\,ks) would result in very compact clumps ($\Delta R\sim2$--$3\,R_{\rm g}$) located far away from the source (at parsec-scales) which appears an implausible scenario. Furthermore, if these compact clumps are instead located in the vicinity of the X-ray source, the ionization of the absorber would reach unrealistically high values. This and the fact that the time-averaged SED from 2007 requires little absorption suggests that the short-term variability is unlikely to be caused by variable absorbing clouds.
\\
\indent The spectral variability is most likely driven by fluctuations in a two-component continuum, which takes the form of soft excess below 1\,keV as well as the hard power-law tail above 2\,keV. During the hard flaring events, in the first half of the observation, the contribution of the soft excess is negligible and the power-law component assumes a relatively hard value ($\Gamma\sim2$). During the latter soft flares, the soft excess increases in flux by at least a factor of 3 while the hard X-ray power-law component appears to soften ($\Gamma\sim2.5$) in response. Overall the variability of \pds may be accounted for by coronal changes, whereby an increase in soft photons from the Comptonized Wien-tail of the disc emission cools the hot coronal electrons, subsequently leading to a steepening of the primary power-law during the soft flaring episodes.

\section{acknowledgements}

We thank the anonymous referee for their careful report, which helped us improving the clarity of the paper. This research has made use of data obtained from the \suzaku satellite, a collaborative mission between the space agencies of Japan (JAXA) and the USA (NASA). GAM and MTC acknowledge support from an STFC studentship while JNR and EN also acknowledge the financial support of STFC via an STFC consolidated grant. JNR also acknowledges support from NASA grant NNX15AF12G. TJT acknowledges support from NASA grant NNX11AJ57G.

\bibliographystyle{mn2e}
\bibliography{gabi_second_paper}

\end{document}